\documentclass[%
 reprint,
showpacs,
 amsmath,amssymb,
 aps,
 pra,
longbibliography,
floatfix,
superscriptaddress
]{revtex4-1}

\usepackage[utf8]{inputenc}
\usepackage[T1]{fontenc}
\usepackage[]{graphicx}
\usepackage{color}
\usepackage[english]{babel}

\usepackage[bookmarks,bookmarksopen,bookmarksdepth=2]{hyperref}
\hypersetup{
    colorlinks=true,
    citecolor=blue,
    linkcolor=blue,
    filecolor=magenta,
    urlcolor=cyan,
}

\usepackage{url}

\usepackage{microtype}
\usepackage{braket}
\usepackage{esvect}
\usepackage[export]{adjustbox}

\usepackage{amsmath,amsfonts,amssymb}
\usepackage{mathtools}
\usepackage{bbm}

\usepackage[position=top,caption=false]{subfig}

\let\originaleqref\eqref
\renewcommand{\eqref}{Eq.~\originaleqref}

\newcommand{\fref}[1]{Fig.~\ref{#1}}

\graphicspath{
{figures/}
}

\begin{document}
\title{Measuring out-of-time-ordered correlation functions without reversing time evolution}

\author{Philip Daniel Blocher}
\email{blocher@unm.edu}
\affiliation{Center for Quantum Information and Control, Department of Physics and Astronomy, University of New Mexico, Albuquerque, New Mexico 87131, USA}
\affiliation{Center for Complex Quantum Systems, Department of Physics and Astronomy, Aarhus University, Ny Munkegade 120, DK-8000 Aarhus C, Denmark}

\author{Serwan Asaad}
\altaffiliation{Currently at Center for Quantum Devices, Niels Bohr Institute, University of Copenhagen, Copenhagen, Denmark.}
\author{Vincent Mourik}
\altaffiliation{Currently at Peter Gr\"unberg Institute, Forschungszentrum J\"ulich GmbH, J\"ulich, Germany}
\author{Mark A. I. Johnson}
\author{Andrea Morello}
\affiliation{School of Electrical Engineering and Telecommunications, UNSW Sydney, Sydney, New South Wales 2052, Australia}

\author{Klaus M{\o}lmer}
\affiliation{Center for Complex Quantum Systems, Department of Physics and Astronomy, Aarhus University, Ny Munkegade 120, DK-8000 Aarhus C, Denmark}

\date{\today}
\bigskip

\begin{abstract} 
Out-of-time-ordered correlation functions (OTOCs) play a crucial role in the study of thermalization, entanglement, and quantum chaos, as they quantify the scrambling of quantum information due to complex interactions.
As a consequence of their out-of-time-ordered nature, OTOCs are difficult to measure experimentally. 
Here we propose an OTOC measurement protocol that does not rely on the reversal of time evolution and is easy to implement in a range of experimental settings. 
The protocol accounts for both pure and mixed initial states, and is applicable to systems that interact with environmental degrees of freedom. 
We demonstrate the application of our protocol by the characterization of scrambling in a periodically-driven spin that exhibits quantum chaos.
\end{abstract}

\maketitle
\noindent

\section{Introduction}
Collective effects in quantum systems play an important role in the development of quantum technologies. 
Quantum entanglement may increase the sensitivity of quantum metrology, and many-body systems scramble initially localized information and self-thermalize. 
Studies of classically chaotic systems in the quantum limit call for suitable measures of quantum chaos, and recent works have shown strong connections between quantum information scrambling, self-thermalization, entanglement generation, and quantum chaotic dynamics \cite{NatCommun.10.1581,NatPhysRev.1.627,PhysRevA.94.040302,Maldacena2016,PhysRevE.100.042201}.

One characterization of the dynamics of quantum information is found in out-of-time-ordered correlation functions (OTOCs), which do not obey the usual time ordering of their constituent operators. 
A particular OTOC is 
\begin{equation}
F(t) = \braket{W^\dagger(t) V^\dagger(0) W(t) V(0)}, \label{eq:OTOC}
\end{equation}
where $V(0)$ and $W(t)$ are Hermitian or unitary operators evaluated in the Heisenberg picture at times $0$ and $t$, respectively. 
Originally introduced in the description of how electron momenta fail to commute at different times in superconductor physics~\cite{Larkin1969}, this particular OTOC has seen a revival as a way of measuring the scrambling of quantum information \cite{NatCommun.10.1581,PhysRevA.99.051803}.

\begin{figure}
\center
\includegraphics[width=1\linewidth]{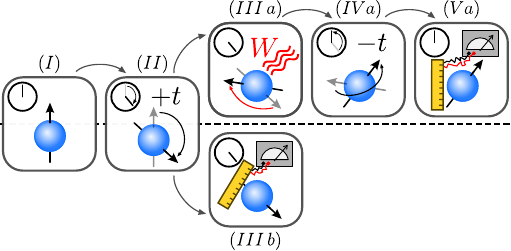}
\caption{Experimental measurement of out-of-time-ordered correlation functions. \textbf{Upper path:} Illustration of the Loschmidt echo procedure applied to a quantum spin. 
(I):~preparation of initial state at time $0$. 
(II):~forward evolution to time $t$. 
(IIIa):~application of perturbation $W$ at time $t$. 
(IVa):~backward evolution to the initial time $0$. 
(Va):~measurement of final state and comparison with the initial state. 
\textbf{Lower path:} The experimental protocol proposed in this article. 
After system initialization~(I) and forward time evolution~(II), we {\it do not} apply any perturbation but instead measure the expectation value of the unitary operator~$W$~(IIIb). 
This yields the OTOC \eqref{eq:OTOC} without requiring the reversal of time evolution.
}
\label{figure:LoschmidtEcho}
\end{figure}

Through the scrambling process, initially localized information -- e.g., a local perturbation -- is spread throughout the particular system's degrees of freedom, becoming inaccessible to any local probes at later times. 
A measure of scrambling may be introduced as the commutator between two operators separated in time,
\begin{equation}
C(t) = \braket{\,\vert [V(0), \, W(t)]\vert^2}. \label{eq:ScramblingCommutator}
\end{equation}
In quantum chaotic systems $C(t)$ may exhibit an exponential growth with a quantum Lyapunov exponent \cite{Maldacena2016,PhysRevE.100.042201}. 
This signals that the later operator $W$ becomes sensitive to the earlier perturbation $V$. 
One may show that $C(t)$ can be related directly to the OTOC $F(t)$ \cite{NatCommun.10.1581,PhysRevA.94.040302,Maldacena2016}. 
In particular, if we let $V(0) = \rho_0 = \ket{\Psi_0}\bra{\Psi_0}$ be the projection operator onto an initial pure system state and assume $W$ to be unitary, we find that $C(t) = 1 - F(t)$. Although $C(t)$ may exhibit exponential growth in quantum chaotic systems, we caution the reader that the presence of exponential scrambling does not necessarily imply quantum chaos, as exponential scrambling may also occur due to the presence of saddle points \cite{PhysRevLett.124.140602,PhysRevA.103.033304}. Methods of distinguishing the different sources of exponential scrambling are discussed in detail in Ref.~\cite{PhysRevA.103.033304}.

Due to the out-of-time-ordered nature of OTOCs, it has proven experimentally challenging to measure the OTOC $F(t)$ as at first glance this requires the reversal of time evolution $0\rightarrow t \rightarrow 0$. 
The reversal of time evolution was implemented experimentally in Ising model quantum simulators by changing the sign of the Hamiltonian $H \rightarrow -H$, thus allowing OTOCs to be measured in a Loschmidt echo-like procedure \cite{Garttner2017,PhysRevLett.120.040402,PhysRevX.7.031011,PhysRevLett.123.090605}. 
This procedure is illustrated in \fref{figure:LoschmidtEcho} for a single spin. 
For systems that do not permit time reversal, the measurement of $F(t)$ was proposed and experimentally demonstrated using interferometric approaches \cite{arXiv.1607.01801}, auxiliary degrees of freedom \cite{nature.567.61}, or statistical correlations \cite{PhysRevX.9.021061,PhysRevLett.124.240505}. 
There are, however, experimental systems for which these methods are insufficient or unsuitable.

In this article we propose an experimental protocol that allows the measurement of the OTOC $F(t)$ without using reversal of time evolution, while also constraining ourselves to a single instance of the system. 
This is done by measuring the expectation value of the operator $W$ at time $t$, $\braket{W(t)}$, as we show below that $F(t) = \vert \braket{W(t)} \vert^2$ under the constraint that the operator $V$ is chosen as the projection onto the initial state of the system. The protocol is thus as visualized in \fref{figure:LoschmidtEcho}(b). 
Our protocol applies to systems admitting an effective spin description, and we demonstrate the protocol's use in characterizing scrambling in a quantum chaotic system: the quantum driven top. 
A single spin-7/2 donor in a solid state device has recently been proposed to realize the quantum driven top~\cite{PhysRevE.98.042206}, thus allowing the exploration of quantum scrambling in small, closed quantum systems.

The structure of this article is as follows: In Section~\ref{section:Protocol} we briefly recall how protocols based on the Loschmidt echo may access OTOCs in quantum systems, and we subsequently propose a novel protocol for measuring OTOCs without using time reversal. 
Our protocol is presented for systems admitting unitary time evolution of a pure initial system state, and subsequently we demonstrate the validity of the protocol in the case of open quantum systems and mixed initial states. 
In Section~\ref{section:Hamiltonian} we briefly discuss the properties and characteristics of the classical and quantum driven top, followed by Sec.~\ref{section:MainResults} where we highlight the application of our OTOC measurement protocol through the example of quantum information scrambling in a driven nuclear spin realizing the quantum driven top.
Finally, in Sec.~\ref{section:Outlook} we provide a discussion and outlook. Throughout this article we let $\hbar = 1$.

\section{Protocol}\label{section:Protocol}
We motivate our protocol by briefly considering the Loschmidt echo, which is the quintessential experiment for probing the sensitivity of quantum time evolution to perturbations \cite{GORIN200633}. The Loschmidt echo is visualized in \fref{figure:LoschmidtEcho} for a single spin: a pure initial state, given by the density matrix $\rho_0 \equiv \rho(0)$, is first evolved forward in time to $t > 0$. A unitary perturbation $W$ (e.g., a rotation) is then applied to the system state at time $t$. Finally, the system is evolved backward in time to $t = 0$. The resulting state is
\begin{equation}
\rho_W(0;t) = W(t)\, \rho_0\, W^\dagger(t), \label{eq:LoschmidtFinalState}
\end{equation}
where the Heisenberg picture operator $W(t) = \mathcal{U}^\dagger(t,0)\, W\, \mathcal{U}(t,0)$ captures the Loschmidt echo procedure described above. In the Loschmidt echo we measure the fidelity $L(t) = \text{Tr}[\rho_0\, \rho_W(0;t)]$.
This fidelity provides a measure of how similar our final state is to the initial state, and consequently how susceptible the system dynamics are to the perturbation $W$ applied at time $t$.

The Loschmidt echo fidelity $L(t)$ is an OTOC on the form of \eqref{eq:OTOC} \cite{Garttner2017,PhysRevLett.120.040402,NatCommun.10.1581}: Choosing one operator $V$ to be the initial state, $V(0) = \rho_0$, assuming an initially pure state (such that $\rho_0 = \ket{\Psi_0}\bra{\Psi_0}$), and requiring $W$ to be unitary, we find that (see appendix~\ref{App:LE-OTOC} for details)
\begin{equation}
L(t) = \text{Tr}[\rho_0\, W^\dagger(t)\, V^\dagger(0)\, W(t)\, V(0)] = F(t). \label{eq:OTOCLoschmidtLink}
\end{equation}
The Loschmidt echo thus provides direct experimental access to $F(t)$, provided that one can implement an experimental protocol that measures the echo fidelity $L(t) = \text{Tr}[\rho_0\, \rho_W(0;t)]$ \cite{Garttner2017}.

We now define our experimental protocol that, contrary to the Loschmidt echo, allows the measurement of $F(t)$ without requiring the use of reversal of time evolution. We assume the operator $V$ to be the projection operator onto the initial pure state, $V(0) = \rho_0 = \ket{\Psi_0}\bra{\Psi_0}$, with no restrictions on the operator $W$.
Inserting this choice of $V$ into \eqref{eq:OTOC} yields
\begin{align}
    F(t) =& \braket{W^\dagger(t)\, V^\dagger(0)\, W(t)\, V(0)} \nonumber\\
    =& \text{Tr}[\rho_0\, W^\dagger(t)\, \rho_0\, W(t)\, \rho_0], 
\end{align}
and carrying out the trace then yields
\begin{align}
    F(t) =& \bra{\Psi_0}W^\dagger(t) \ket{\Psi_0}\bra{\Psi_0} W(t)\ket{\Psi_0} \nonumber\\
    \equiv& \braket{W^\dagger(t)} \braket{W(t)} \nonumber\\
    =& \vert \braket{W(t)} \vert^2. \label{eq:UnravelingOTOC}
\end{align}
Here the expectation value is with respect to the initial state $\ket{\Psi_0}$. 
Equation~(\ref{eq:UnravelingOTOC}) reveals that $F(t)$ may be determined experimentally by measuring either $\braket{W(t)}$ or $\vert \braket{W(t)} \vert^2$ directly from the time-evolved state $\rho(t)$ at time $t$ as visualized in \fref{figure:LoschmidtEcho} step $(IIIb)$. We thus obtain an OTOC measurement protocol that does not require the reversal of time evolution. The protocol is valid for any operator $W$, in particular for both Hermitian observables and unitary perturbations.

The measurement of the expectation value $\braket{W(t)}$ can be broken down into two relevant cases. For a Hermitian observable $W$, the expectation value $\braket{W(t)}$ can be directly measured in the experiment.
If instead $W$ is unitary, we may write $W$ as the complex exponential of a (Hermitian) perturbation Hamiltonian $H^\prime$, $W = \exp(-i H^\prime \Delta t)$.
While we cannot directly measure $\braket{W(t)}$ when $W$ is not Hermitian, we show in the following how it can be reconstructed by measuring $H^\prime$ at time $t$.
The eigenstates and eigenvalues of $H^\prime$ follow from the spectral theorem: $H^\prime \ket{\psi_n} = E_n^\prime \ket{\psi_n}$.
The eigenstates of $W$ are identical to those of $H^\prime$ with eigenvalues $\mu_n$: $W \ket{\psi_n} = \exp(-i E_n^\prime \Delta t) \ket{\psi_n} \equiv \mu_n \ket{\psi_n}$.
By sampling the time-evolved state $\ket{\Psi(t)} = \mathcal{U}(t,0)\ket{\Psi_0}$ in the eigenbasis of $H^\prime$ (and thus also $W$), we obtain the state populations $\vert c_n(t) \vert^2$, where $c_n(t) = \braket{\psi_n \vert \Psi(t)}$ is the time-dependent amplitude of the $n$th eigenstate. 
With these populations we can reconstruct $\braket{W(t)}$ by weighting the eigenvalues of $W$ with the corresponding state population:
\begin{equation}
    \braket{W(t)} = \sum_n \vert c_n(t) \vert^2 \mu_n.\label{eq:expectationvalueexpanded}
\end{equation}

The experimental overhead for measuring the unitary operator $W = \exp(-i H^\prime \Delta t)$ is the sampling required to obtain the eigenstate populations of $H^\prime$, where $H^\prime$ is suitably chosen for the system of interest so as to have experimentally accessible eigenstate populations. In particular, the chosen observable $H^\prime$ should have known eigenvalues and eigenvectors. The choice of $H^\prime$ is exemplified in section~\ref{subsection:ChoosingW} for the quantum driven top.

By varying $\Delta t$ we change the eigenvalues $\mu_n$ while leaving the eigenstates $\ket{\psi_n}$ (and thus also the state amplitudes $c_n(t)$) unchanged. Hence, we may obtain the OTOC $F(t)$ for operators $W$ corresponding to different $\Delta t$ after having measured the state populations $\vert c_n(t)\vert^2$ only once.

A brief comparison between our scheme and the Loschmidt echo protocol shows that both provide experimental access to the OTOC $F(t)$ using the same assumptions on the operators $V$ and $W$. However, the requirements for the experimental system as well as the role of the operator $W$ in the experiment differ significantly. In the Loschmidt echo protocol, a perturbation represented by the unitary operator $W$ must be applied to the system state at time $t$, and the reversal of time evolution is necessary to compare the perturbed state to the initial state at time $0$. In our protocol, $W$ is not applied as a perturbation but the populations of its eigenstates must be measured at time $t$, and the inferred value of $\vert \braket{W(t)}\vert^2$ then yields the desired OTOC $F(t)$.

In the above description of the protocol we have considered the unitary time evolution of a pure initial state for simplicity, and retained the choice of $V(0) = \rho_0$ from the Loschmidt echo. In the following section~\ref{subsection:mixed} we demonstrate that the protocol also applies for mixed initial states, and that we may extend the protocol to arbitrary choices of the operator $V$. In section~\ref{subsection:OpenQuantumSystem} we demonstrate that our protocol is able to approximate the OTOC for open quantum systems that interact weakly with their environment, and we validate the use of the protocol in the presence of decoherence and dissipation by evaluating the environmental contributions using the methods of Ref.~\cite{PhysRevA.99.033816}.

\subsection{Extension to mixed initial states and arbitrary operators}\label{subsection:mixed}
The OTOC measurement protocol is readily extended to the case of a mixed initial state $\rho_0 = \sum_n p_n \rho_n$, where $\rho_n = \ket{\Psi_n}\bra{\Psi_n}$ is a pure system state, the states $\ket{\Psi_n}$ form an orthonormal basis, and the statistical weights satisfy $0 \leq p_n < 1$, $\sum_n p_n = 1$. 
For now we retain the choice of $V(0) = \rho_0$, though later we show that this restriction may be relaxed.  
By insertion of $\rho_0$ and $V(0) = \rho_0$ into \eqref{eq:OTOC} it follows that
\begin{align}
F(t) =& \sum_{l,m,n} p_l\,p_m\,p_n\,\text{Tr}\Big[ \ket{\Psi_l}\bra{\Psi_l} W^\dagger(t)\, \Pi_m\, W(t)\, \Pi_n \Big] \nonumber\\
=& \sum_{m,n} p_n^2 p_m \braket{W^\dagger(t)\, \Pi_m\, W(t)}_n, \label{eq:MixedOTOC}
\end{align}
where $\Pi_m = \ket{\Psi_m}\bra{\Psi_m}$ and $\braket{\cdots}_n = \text{Tr}[\rho_n \cdots]$ denotes the expectation value with respect to the pure initial state $\rho_n$. 
According to \eqref{eq:MixedOTOC} the OTOC $F(t)$ may be evaluated by an appropriately weighted sum over the expectation values $\braket{W^\dagger(t)\, \Pi_m\, W(t)}_n$.

Let us consider a single term, $\braket{W^\dagger(t) \Pi_m W(t)}_n$ and apply the explicit expression 
$\ket{\Psi_{m(n)}}\bra{\Psi_{m(n)}}$ for $\Pi_m$ ($\rho_n$):
\begin{align}
\braket{W^\dagger(t)\, \Pi_m\, W(t)}_n =&\text{Tr}[\rho_n W^\dagger(t) \Pi_m W(t)] \nonumber\\
=& \bra{\Psi_n} W^\dagger(t) \ket{\Psi_m} \bra{\Psi_m} W(t) \ket{\Psi_n} \nonumber\\
=& \vert \bra{\Psi_m} W(t) \ket{\Psi_n} \vert^2. \label{eq:MixedOTOCElement}
\end{align}
In the second equality we have used that $\Pi_m$ is a rank one projector to separate the trace into the product of two matrix elements. 
To evaluate $\bra{\Psi_m}W(t)\ket{\Psi_n}$ experimentally we first define the states
\begin{align}
\ket{a^\pm_{nm}} =& \frac{1}{\sqrt{2}}\big(\ket{\Psi_n} \pm\ket{\Psi_m}\big), \label{eq:SuitableStateA}\\
\ket{b^\pm_{nm}} =& \frac{1}{\sqrt{2}}\big(\ket{\Psi_n} \pm i \ket{\Psi_m}\big), \label{eq:SuitableStateB}
\end{align}
and it then follows from the polarization identity that
\begin{widetext}
\begin{equation}
2\bra{\Psi_m}W(t)\ket{\Psi_n} = \bra{a^+_{nm}}W(t)\ket{a^+_{nm}} -  \bra{a^-_{nm}}W(t)\ket{a^-_{nm}} + i\bra{b^+_{nm}}W(t)\ket{b^+_{nm}} - i  \bra{b^-_{nm}}W(t)\ket{b^-_{nm}}. \label{eq:PolarizationIdentity}
\end{equation}
\end{widetext}
This allows us to access the matrix elements $\bra{\Psi_m}W(t)\ket{\Psi_n}$ and $\bra{\Psi_n} W(t) \ket{\Psi_m}$ by experimentally measuring the four expectation values $\bra{a^\pm_{nm}}W(t)\ket{a^\pm_{nm}}$, $\bra{b^\pm_{nm}}W(t)\ket{b^\pm_{nm}}$. Each expectation value may be evaluated experimentally in the same manner as discussed for \eqref{eq:expectationvalueexpanded} by measuring the operator $W$ in a forward-only time evolved state. Using \eqref{eq:MixedOTOC}, the OTOC $F(t)$ follows from weighted sums over these expectation values. The OTOC $F(t)$ for mixed initial states can thus be obtained by our experimental protocol applied to the suitable set of pure states given by Eqs.~(\ref{eq:SuitableStateA}-\ref{eq:SuitableStateB}).

Finally we note that we may relax the restrictions on the operator $V$, which we previously assumed to be the projection onto the initial system state $V(0) = \rho_0$. In fact, we may choose an arbitrary operator 
\begin{equation}
V(0) = \sum_{i,j} v_{ij} \ket{\Psi_i}\bra{\Psi_j},
\end{equation}
where we have written the operator $V(0)$ in the orthonormal basis $\{\Psi_n\}$ introduced previously for the density matrix $\rho_0$ using the complex coefficients $v_{ij}$. Inserting the general operator $V(0)$ into \eqref{eq:OTOC} yields
\begin{align}
F(t) =& \sum_{i,j} \sum_{k,l} v_{ij}\, v_{kl}^\ast\, p_j \bra{\Psi_j} W^\dagger(t) \ket{\Psi_l} \nonumber\\
&\times \bra{\Psi_k} W(t) \ket{\Psi_i}. \label{eq:OTOCArbitraryV}
\end{align}
We note that $\bra{\Psi_j} W^\dagger(t) \ket{\Psi_l} = \bra{\Psi_l} W(t) \ket{\Psi_j}^\ast$. To obtain the OTOC $F(t)$ in \eqref{eq:OTOCArbitraryV}, it is therefore sufficient to evaluate $\bra{\Psi_k} W(t) \ket{\Psi_i}$ via the polarization identity \eqref{eq:PolarizationIdentity}. If $d$ is the dimension of the system's Hilbert space, the evaluation of at most $2 d^2$ expectation values is required to yield the OTOC.

\subsection{OTOCs in open quantum systems}\label{subsection:OpenQuantumSystem}
Previously we derived our OTOC measurement protocol for the case of a closed quantum system undergoing unitary time evolution. Here we highlight the challenging nature of extending OTOC measurement protocols to the dynamics of open quantum systems. 

First, we recall Ref.~\cite{PhysRevA.99.033816} which provides a generalization of the quantum regression theorem allowing us to calculate OTOCs for open quantum systems coupled to a Markovian bath. The OTOC $F(t) = \braket{W^\dagger(t)\, V^\dagger(0)\, W(t)\, V(0)}$ may be expanded as
\begin{equation}
    F(t) = \sum_{i,j}\sum_{m,n} (W^\dagger)_{ji}\, W_{nm}\, \rho_{ij,mn}(t),
\end{equation}
where we have defined the object 
\begin{equation}
    \rho_{ij,mn} := \braket{(\ket{j}\bra{i})(t)\, V^\dagger(0)\, (\ket{n}\bra{m})(t)\, V(0)}
\end{equation}
by introducing the dyadic products $(\ket{j}\bra{i})(t)$. We note that the entries of the density matrix read $\rho_{ij}(t) = \braket{(\ket{j}\bra{i})(t)}$. As in the protocol presented previously in \eqref{eq:UnravelingOTOC}, we let the operator $V(0) = \rho(0) = \ket{\Psi_0}\bra{\Psi_0}$ be the pure density matrix at time $t = 0$. With this particular choice of $V(0)$, at time $t = 0$ the object $\rho_{ij,mn}$ is a simple product of density matrix elements,
\begin{align}
    \rho_{ij,mn}(0) =& \braket{\ket{j}\bra{i}\rho(0)\ket{n}\bra{m} \rho(0)} \nonumber\\
    =& \braket{i \vert \Psi_0} \braket{\Psi_0 \vert n} \braket{m \vert \Psi_0} \braket{\Psi_0 \vert j} \nonumber\\
    =& \rho_{ij}(0)\, \rho_{mn}(0).
\end{align}
Writing the density matrix master equation on the form
\begin{equation}
    \dot{\rho}_{ij}(t) = \sum_{i^\prime, j^\prime} M_{ij,i^\prime j^\prime}\, \rho_{i^\prime j^\prime}, \label{eq:WrongOTOCTimeEvo}
\end{equation}
where $\dot{Q} \equiv \frac{\mathrm{d}Q}{\mathrm{d}t}$,
we find that the time evolution of the object $\rho_{ij,mn}$ reads~\cite{PhysRevA.99.033816}
\begin{align}
    \dot{\rho}_{ij,mn}(t) =& \sum_{i^\prime, j^\prime} M_{ij,i^\prime j^\prime}\, \rho_{i^\prime j^\prime ,mn}(t) \nonumber\\
    &+ \sum_{m^\prime, n^\prime} M_{mn,m^\prime n^\prime}\, \rho_{i j ,m^\prime n^\prime}(t) \nonumber\\
    &+ \text{environment contributions}. \label{eq:noiseterms}
\end{align}
The environment contribution terms in the last line stem from the bath operators that contribute to the Heisenberg picture time evolution of system observables. These terms have vanishing mean and do not appear in the master equation \eqref{eq:WrongOTOCTimeEvo}. Under the Markov assumption, the bath operators have no correlations with the system observables at previous times, and hence they do not contribute to time-ordered correlation functions. However, the system observables retain correlations from their previous interactions with the bath observables, and in the time-evolution of out-of-time-ordered objects like \eqref{eq:noiseterms}, the product with environment operators at the same earlier times causes the deterministic environment terms in \eqref{eq:noiseterms}. Their precise nature is discussed in detail in Ref.~\cite{PhysRevA.99.033816} where, in specific studies with system-environment interactions of modest strength, these terms were found to contribute only negligibly to the OTOC. In practical cases one may thus neglect the environment contributions in \eqref{eq:noiseterms}.

Let us now consider the outcome of employing our measurement protocol -- derived exclusively for a closed system in \eqref{eq:UnravelingOTOC} -- in an open quantum system. By merely evaluating and multiplying the measured expectation values of the operator $W(t)$ in the open system, we get the result
\begin{equation}
    \widetilde{F}(t) := \braket{W^\dagger(t)}\braket{W(t)} = \sum_{i,j} \sum_{m,n} (W^\dagger)_{ji}\, W_{nm}\, \widetilde{\rho}_{ij,mn}(t)
\end{equation}
where we have defined $\widetilde{\rho}_{ij,mn}(t) := \rho_{ij}(t)\, \rho_{mn}(t)$. Since $\widetilde{\rho}_{ij,mn}(t)$ obeys the equation of motion
\begin{align}
    \dot{\widetilde{\rho}}_{ij,mn}(t) =& \sum_{i^\prime, j^\prime} M_{ij,i^\prime j^\prime}\, \widetilde{\rho}_{i^\prime j^\prime ,mn}(t) \nonumber\\
    &+ \sum_{m^\prime, n^\prime} M_{mn,m^\prime n^\prime}\, \widetilde{\rho}_{i j ,m^\prime n^\prime}(t), \label{eq:wrongEoM}
\end{align}
and at the initial time reads $\widetilde{\rho}_{ij,mn}(0) = \rho_{ij}(0) \rho_{mn}(0) \equiv \rho_{ij,mn}(0)$, it follows that $\widetilde{\rho}_{ij,mn}(t) = \rho_{ij,mn}(t)$ (and thus $\widetilde{F}(t) = F(t)$) for all times $t$ if one disregards the noise terms in \eqref{eq:noiseterms}. The simple experimental protocol proposed in \eqref{eq:UnravelingOTOC} thus provides access to the exact same approximation to the OTOC as we obtain by theoretically disregarding the environment contributions in \eqref{eq:noiseterms}. Through the method presented in Ref.~\cite{PhysRevA.99.033816}, one may evaluate the importance of the noise correlation terms on a case-by-case basis and thus validate the use of the closed system protocol presented in this paper to experimentally determine the values of OTOCs for open quantum systems.

The role of correlations between the bath and system operators can also be understood by 
considering the unitary dynamics of the combined system + environment from an initial product state $\ket{\Psi_{SE}(t=0)} = \ket{\Psi_S(0)}\otimes\ket{\Psi_E(0)}$, where the subscript $S$ ($E$) denotes the system (environment). We write the system operator $V(0)$ as $V(0) = V_S(0) \otimes \mathbbm{1}_E$, and assume that $V_S$ is the projector onto the initial system state, $V_S(0) = \ket{\Psi_S(0)}\bra{\Psi_S(0)}$. The identity operator for the environment may be expanded as $\mathbbm{1}_E(0) = \sum_\alpha \ket{\alpha_E(0)}\bra{\alpha_E(0)}$, and inserting into \eqref{eq:OTOC} then yields
\begin{equation}
F(t) = \sum_\alpha \big\vert \bra{\Psi_S(0)}\bra{\alpha_E(0)} W(t) \ket{\Psi_S(0)}\ket{\Psi_E(0)} \big\vert ^2, \label{eq:OTOCDifferentEnvironments}
\end{equation}
where we note that the operator $W(t)$ is evaluated between different environment states $\ket{\alpha_E}$ and $\ket{\Psi_E(0)}$. At the initial time $t=0$, the operator $W(0) := W_S(0) \otimes \mathbbm{1}_E$ acts like a system operator. However, for non-vanishing times $t > 0$ the system-environment interaction causes $W$ to take on mixed characteristics of the system and environment operators. Although in principle \eqref{eq:OTOCDifferentEnvironments} may be evaluated using the polarization identity, one would need to consider all environmental states to construct the exact OTOC from measurements. Discarding the noise contributions in \eqref{eq:noiseterms} is equivalent to discarding the mixing of system and environment degrees of freedom in $W(t)$, and when the former approximation can be justified, the OTOC measurement protocol holds approximately for open quantum systems.

Finally we note that the presence of non-vanishing system-environment interactions such as decoherence may cause the decay of the OTOC signal and other correlation functions to show false positives when quantifying quantum information scrambling in the system \cite{PhysRevLett.129.050602}. This is important for the near-term implementation of proposed experimental OTOC measurement protocols on noisy intermediate scale quantum (NISQ) simulators and devices. Reassuringly, even in the presence of decoherence, our proposed OTOC measurement protocol approximately captures the OTOC dynamics up to a small error caused by the omission of the noise terms in the equations of motion. Separate measures may then be taken to distinguish the desired quantum information scrambling signal from the signal due to environmental couplings (e.g. Ref.~\cite{PhysRevLett.129.050602}).

\section{The classical and quantum driven top}\label{section:Hamiltonian}
Our novel protocol described above provides an effective way to calculate the OTOC of a system without reversing its time evolution.
This enables the study of quantum information scrambling in a large range of systems for which existing OTOC protocols are incompatible.
To demonstrate its efficacy, we apply our protocol to a realistic experimental proposal of the quantum-chaotic driven top system~\cite{haake1990optical}, a variant of the kicked top~\cite{haake1987classical,scharf1988kramers,zyczkowski1990indicators,kus1991quantum,schack1994hypersensitivity,jacquod2001golden}, implemented using the large nuclear spin of a single group-V donor in silicon \cite{PhysRevE.98.042206}. 
Prior to investigating scrambling within the quantum driven top, it is instructive to consider the classical chaos present in its classical analogue. In this section we provide a brief overview of both the classical and quantum driven top, as well as motivate the choice of suitable initial states and operator $W$ for our OTOC measurement protocol.

\subsection{Chaos in the classical driven top}\label{subsection:ClassicalChaos}
In a classical system exhibiting chaos, an initial perturbation $\delta x(t=0)$ is amplified exponentially over time as $\delta x(t) \approx e^{\lambda t} \delta x(0)$, where $\lambda$ is the average Lyapunov coefficient (see App.~\ref{App:VariationalEquation} for estimating the Lyapunov coefficient).

The classical driven top is described by the Hamiltonian
\begin{equation}
\mathcal{H} = \alpha L_z + \frac{\beta}{\vert \mathbf{L} \vert} L_x^2 + \gamma \cos(\omega t) L_y, \label{eq:ClassicalH}
\end{equation}
where $L_i$ is the angular momentum component in the $i$th direction, $\alpha$ and $\beta$ are constants, and $\gamma$ is the strength of the linear drive with angular frequency $\omega$. The quadratic term has been normalized to the magnitude of the angular momentum $\vert \mathbf{L} \vert$ to ensure that $\alpha = \beta$ implies equal strengths of the static linear and quadratic terms.

\begin{figure}
\center
\includegraphics[width=\linewidth]{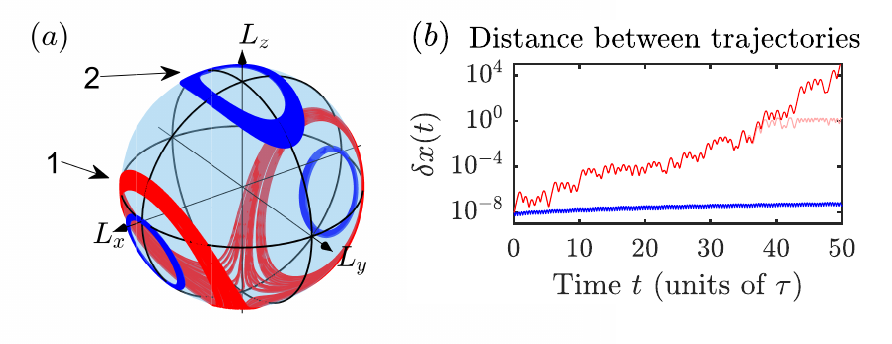}
\caption{Chaotic behavior of a classical driven top. \textbf{(a):}~Classical regular (blue) trajectories and a single chaotic (red) trajectory. Two chaotic lobes enclose islands of regular behavior. \textbf{(b):} Distance $\delta x(t)$ between two adjacent classical regular (blue) and chaotic (pink) trajectories at time $t$ after an initial separation $\delta x(0) = 1e-8$. We have used the system parameters $\beta = 1.5\alpha$, $\gamma = 0.05\alpha$, $\omega = 1.5\alpha$.
The chaotic trajectories diverge exponentially, but reach a plateau at $t \approx 40\tau$ due to the finite size of the phase space.
However, when the variational equation is used to calculate the separation between chaotic trajectories (red), the separation is not limited by the finite system size, and thus provides an accurate estimate of the average Lyapunov exponent (App.~\ref{App:VariationalEquation}).
}\label{figure:ClassicalChaos}
\end{figure}

The choice of coefficients $\alpha$, $\beta$, $\gamma$, and $\omega$ has pronounced effects on the chaoticity of the system's phase space \cite{PhysRevE.98.042206}, which can have coexisting regions of chaotic and regular behavior. These regions may be found by comparing the evolution of initially adjacent trajectories, as illustrated in \fref{figure:ClassicalChaos} for trajectories originating in the regular (blue) and chaotic (red) region.

\begin{figure}
\center
\includegraphics[scale=1]{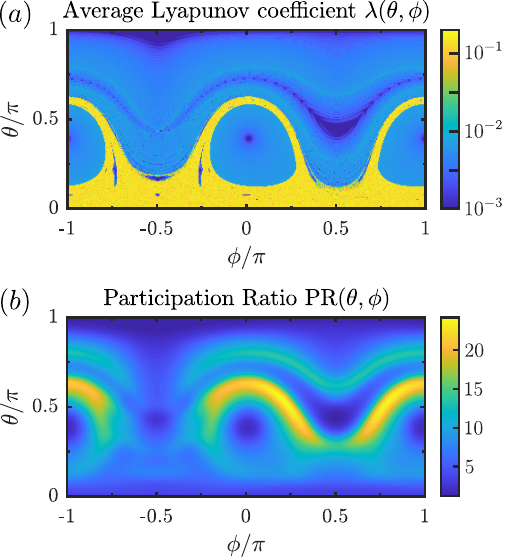}
\caption{Classical Lyapunov exponents and Floquet analysis for the driven top. \textbf{(a):}~Classical Lyapunov exponents $\lambda(\theta,\phi)$, found by solving the variational equation for a total duration $1000\tau$ (App.~\ref{App:VariationalEquation}). The bright yellow area is the chaotic region, with the two chaotic lobes emerging from the chaotic pole at $\theta \approx 0$. The blue areas are regular regions, where $\lambda(\theta,\phi)$ is cut off at a minimum $10^{-3}$. The parameters $\theta$ and $\phi$ are the polar angle and azimuthal angle of the initial angular momentum vector $\mathbf{L}$, respectively. \textbf{(b):}~Participation ratio (see text) of each SCS $\ket{\theta,\phi}$, for $J = 41/2$. We note the similarity with the classical chaotic pattern in (a).}\label{figure:ClassicalQuantumDynamics}
\end{figure}

In \fref{figure:ClassicalChaos}(a) the red curve visualizes a single trajectory in the chaotic region. We see that the lobes of this trajectory divide the phase space into four distinct regions: A chaotic region, a main regular region, and two regular islands inside the lobes of the chaotic region. The average Lyapunov exponents are extracted from these trajectories by comparing the final divergence of two initially adjacent trajectories (App.~\ref{App:VariationalEquation}). The distance between two adjacent trajectories is displayed in \fref{figure:ClassicalChaos}(b) for initial states originating in the chaotic (red) and regular (blue) region, and the extracted average Lyapunov exponents are visualized in \fref{figure:ClassicalQuantumDynamics}(a) for varying polar angles $\theta$ and azimuthal angles $\phi$ of the initial angular momentum vector $\mathbf{L}$. This further illustrates the existence of these four distinct regions.

\subsection{The quantum driven top}\label{subsection:QuantumHamiltonian}
The quantum equivalent of \eqref{eq:ClassicalH} is the quantum driven top described by the following Hamiltonian
\begin{equation}
H(t) = \alpha J_z + \frac{\beta}{J} J_x^2 + \gamma \cos(\omega t) J_y, \label{eq:QuantumH}
\end{equation}
where the $J_i$ are angular momentum operators obeying commutator relations $[J_i,J_j] = i \epsilon_{ijk} J_k$. As in the classical case, the quadratic term is normalized by the magnitude of the angular momentum $J$ to ensure that setting $\alpha = \beta$ yields equally strong contributions from the static linear and quadratic terms.

The linear drive with angular frequency $\omega$ results in a time-dependent periodic Hamiltonian with period $\tau = 2\pi/\omega$. The Floquet formalism for treating time-dependent periodic Hamiltonians hence lends itself to the study of the system's dynamics \cite{JPhysB.49.013001}. The Floquet operator $\mathcal{U}_\mathcal{F} \equiv \mathcal{U}(\tau,0)$, given by the time-ordered integral
\begin{equation}
\mathcal{U}_\mathcal{F} = \hat{T} e^{-i \int_0^\tau \mathrm{d}t^\prime H(t^\prime)} \approx \prod_{k=0}^N e^{-i H(\frac{k}{N}\tau)\frac{\tau}{N}}, \label{eq:FloquetOperatorAndApproximation}
\end{equation}
obeys the property $\ket{\psi(n\tau)} = \mathcal{U}_\mathcal{F}^n \ket{\psi(0)}$ for all states $\ket{\psi(0)}$. Here $\hat{T}$ is the time-ordering operator, and the terms in the right-most expression of \eqref{eq:FloquetOperatorAndApproximation} are likewise multiplied in a time-ordered manner. $N$ is a large number chosen such that $\mathcal{U}_\mathcal{F}$ does not change appreciably by increasing the number of segments $N$.

We will use the following system parameters for the remainder of this article: $\beta = 1.5\alpha$, $\gamma = 0.05\alpha$, and $\omega = 1.5\alpha$. This choice of parameters ensures a mixed classical phase space in which chaotic and regular behavior co-exist, as seen in section~\ref{subsection:ClassicalChaos}.

\subsection{Choosing suitable initial states and perturbation $W$ for the measurement protocol}\label{subsection:ChoosingW}
While our protocol in section~\ref{section:Protocol} for measuring $F(t)$ is applicable to any unitary operator $W$, we may, on a per system basis, choose a suitable $W$ that eases the implementation of the measurement protocol. For quantum chaos purposes, the choice of operators $V$ and $W$ generally has little influence on the universal characteristics of the OTOC \cite{PhysRevLett.124.160603}. This is in the spirit of quantum chaoticity being an intrinsic property of the Hamiltonian rather than a consequence of applied perturbations, and it is thus reassuring that in our protocol the perturbation does not need to be applied; it is merely a quantity that characterizes the evolution of the system under the chaotic Hamiltonian dynamics.

The natural choice for an initial state in a spin system is a spin coherent state (SCS) $\ket{\theta,\phi}$~\cite{PhysRevA.6.2211}.
Spin coherent states have the highest resemblance to classical phase-space states: their spins are maximally aligned along polar (azimuthal) angle $\theta$ ($\phi$) and have a minimal uncertainty along the transverse axes. We therefore let $V = \rho_0 = \ket{\theta,\phi}\bra{\theta,\phi}$ for $\theta \in [0,\pi]$, $\phi \in [0,2\pi]$ for the remainder of this article. The SCS $\ket{\theta,\phi}$ may be obtained by rotating the $J_z$ eigenstate $\ket{J,\,m_J = J}$ first by a polar angle $\theta$ about the $y$-axis, followed by an azimuthal angle $\phi$ about the $z$-axis. These two rotations may be expressed as a single rotation $\mathcal{R}(\theta,\phi)$ by angle $\theta$ about an axis $(-\sin(\phi),\cos(\phi),0)$ in the $xy$-plane, yielding the expression
\begin{align}
\ket{\theta,\phi} =& \mathcal{R}(\theta,\phi) \ket{J,J} \nonumber\\
=& e^{-i \theta (-\sin(\phi) J_x + \cos(\phi) J_y)} \ket{J,J}. \label{eq:SpinCoherentState}
\end{align}

To capture how initially commuting operators fail to commute at later times, we require the operators $V = \ket{\theta,\phi}\bra{\theta,\phi}$ and $W$ to commute at time $0$ so that $C(0)=0$ (see \eqref{eq:ScramblingCommutator}). As the SCS $\ket{\theta,\phi}$ is oriented along the axis $\Omega = (\theta,\phi)$, we choose $W\equiv W_\epsilon (\Omega)$ to be the rotation about the axis $\Omega$ by an angle $\epsilon$, ensuring that the two operators commute at initial times. $W_\epsilon(\Omega)$ is thus given by the expression
\begin{equation}
W_\epsilon(\Omega) = e^{-i \, \epsilon \,\textbf{n}(\Omega)\cdot \textbf{J}}, \label{eq:OTOCRotation}
\end{equation}
where the axis of rotation is given as
\begin{equation}
\textbf{n}(\Omega) = 
\begin{pmatrix}
\sin(\theta)\cos(\phi) \\
\sin(\theta)\sin(\phi) \\
\cos(\theta)
\end{pmatrix},
\end{equation}
and where $\textbf{J} = (J_x, J_y, J_z)^T$ is the vector of angular momentum operators. We note that $W_\epsilon(\Omega)$ may be expressed in terms of elementary rotations~as~\cite{Sakurai2011}
\begin{equation}
W_\epsilon(\Omega) = R_W(\theta,\phi)\, e^{-i J_z \epsilon}\, R_W^\dagger(\theta,\phi),
\end{equation}
where
\begin{equation}
R_W(\theta,\phi) = e^{-i\,\theta (-\sin(\phi)\, J_x + \cos(\phi)\, J_y)} \equiv \mathcal{R}(\theta,\phi)
\end{equation}
is a rotation about the axis $(-\sin(\phi),\, \cos(\phi),\, 0)^T$ by the angle $\theta$ and thus identical to the SCS rotation $\mathcal{R}(\theta,\phi)$ introduced above. Hence, if $\ket{J,m}$ is the $m$th eigenstate of $J_z$ with eigenvalue $m$, the $m$th eigenstate of $W$ with eigenvalues $\mu_m = \exp(-i\,m\,\epsilon)$ is $\ket{\psi_m} = \mathcal{R}(\theta,\phi) \ket{J,m}$.

With the eigenstates $\ket{\psi_m}$ of $W_\epsilon(\Omega)$ found, the populations $\vert c_m\vert^2 = \vert\bra{J,m} \mathcal{R}^\dagger(\theta,\phi) \ket{\Psi(t)}\vert^2$ may be obtained from the following experiment: 
\begin{enumerate}
    \item Prepare the initial spin coherent state $\ket{\Psi_0} = \ket{\theta, \phi}$ at time $0$ and let it evolve forward in time to time $t$, where the state reads $\ket{\Psi(t)}$.
    \item Apply the rotation $\mathcal{R}^\dagger(\theta,\phi)$ to $\ket{\Psi(t)}$ and project the resulting state onto the $J_z$ eigenbasis to sample $\vert c_m\vert^2$.
    \item Repeat steps 1-2 until the desired accuracy of the $\vert c_m\vert^2$-coefficients is obtained.
\end{enumerate}
The desired OTOC $F(t)$ then follows from Eqs.~(\ref{eq:UnravelingOTOC})-(\ref{eq:expectationvalueexpanded}).

Analogous to the interaction time $\Delta t$ in our introduction of the OTOC measurement protocol in section~\ref{section:Protocol}, the rotation angle $\epsilon$ provides tunability of the unitary rotation operator and hence its eigenvalues $\mu_m$, while leaving the eigenstates $\ket{\psi_m}$ unchanged. This allows us to study OTOCs for a family of unitary operators (rotations by varying angles around the axis $\Omega$) after having conducted only a single experiment.

\subsection{Floquet component analysis}\label{subsection:FloquetAnalysis}
Finally, prior to investigating the OTOC behavior for the quantum driven top in section~\ref{section:MainResults}, it is illuminating to consider the system dynamics induced by the Hamiltonian during time evolution. As described above, the Hamiltonian (given in \eqref{eq:QuantumH}) is periodic with period $\tau = 2\pi / \omega$. The system's time evolution for a single period $\tau$ is therefore described by the Floquet operator $\mathcal{U}_\mathcal{F}$ (see \eqref{eq:FloquetOperatorAndApproximation}), and the eigenvalues and eigenstates of the Floquet operator may provide insight into the system dynamics. As $\mathcal{U}_\mathcal{F}$ is unitary, we may write its eigenvalues as $f_i = \exp(-i \omega_i \tau)$, in accordance with Floquet's theorem. Analogous to the usual time evolution operator $\mathcal{U}(t)$, the pseudoeigenfrequencies $\omega_i$ set the timescales of the system dynamics.

The system dynamics may be visualized by expanding each SCS $\ket{\theta,\phi}$ on the Floquet eigenstates and considering the expansion amplitudes. We here use the inverse participation ratio (IPR) \cite{npjQuantInf.5.78} as a measure of whether the SCS $\ket{\theta,\phi}$ is dominated by a single or few Floquet eigenstates (thus only a few expansion amplitudes will be significant) or instead delocalized in the Floquet eigenbasis (with many non-vanishing expansion amplitudes). Letting $\ket{\omega_i}$ denote the $i$th Floquet eigenstate, the IPR for the SCS $\ket{\theta,\phi}$ is given as
\begin{equation}
\text{IPR}(\theta,\phi) = \sum_{i=1}^d \vert \braket{\theta,\phi\,\vert\, \omega_i} \vert^4,
\end{equation}
where $d$ is the dimension of the Hilbert space. The participation ratio (PR) follows as $\text{PR}(\theta,\phi) = \text{IPR}(\theta,\phi)^{-1}$. For SCS dominated by a few Floquet eigenstates, $\text{PR}(\theta,\phi)$ will be close to unity, with unity only if $\ket{\theta,\phi} \equiv \ket{\omega_k}$ for some index $k$. For SCS delocalized in the Floquet eigenbasis, thus being a complex superposition of many Floquet eigenstates, $\text{PR}(\theta,\phi)$ will instead tend toward $N$.

In \fref{figure:ClassicalQuantumDynamics}(b) we visualize the system dynamics using the PR for $J = 41/2$. We observe by comparison with \fref{figure:ClassicalQuantumDynamics}(a) that SCS corresponding to classically regular regions in general have low PRs and are comprised of few Floquet eigenstates, while SCS in classically chaotic regions have high PRs and are comprised of many significant Floquet eigenstates, thus accessing a much larger portion of the Hilbert space.

\section{OTOC dynamics in parameter space}\label{section:MainResults}
We now demonstrate the application of our OTOC measurement protocol presented in section~\ref{section:Protocol} to calculate the OTOC $F(t)$ for the quantum driven top with the spin coherent states $\{\ket{\theta,\phi}\}$ as the initial states. We remind ourselves of the choice of operators $V = \rho_0 = \ket{\theta,\phi}\bra{\theta,\phi}$ and $W\equiv W_\epsilon(\Omega)$ (as given by \eqref{eq:OTOCRotation}) for each $\theta \in [0,\pi]$ and $\phi \in [0, 2\pi]$.

\begin{figure*}
\center
\includegraphics[width=1\linewidth]{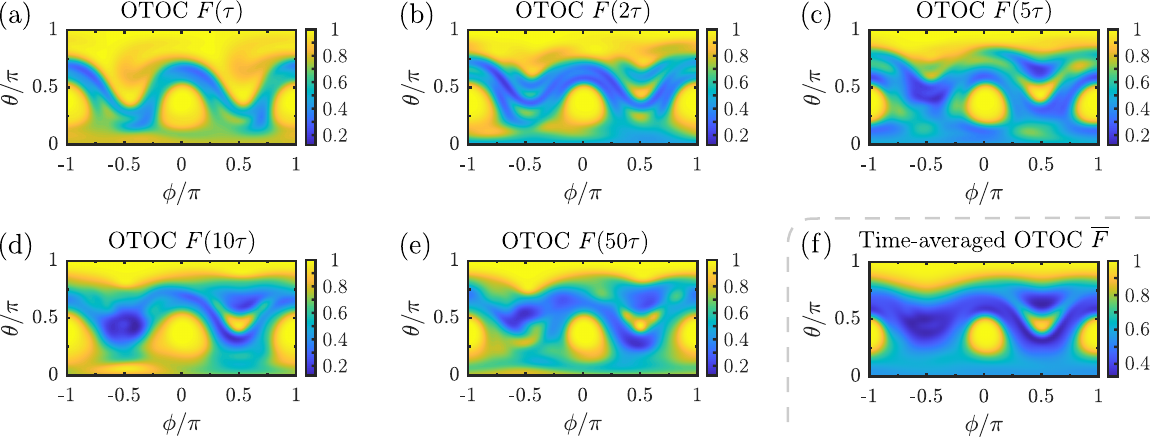}
\caption{OTOC $F(t)$ for spin $J = 41/2$ and rotation angle $\epsilon = \pi/40$. \textbf{(a-e)}: Snapshots of $F(t)$ at $t = 1\tau$, $2\tau$, $5\tau$, $10\tau$, and $50\tau$. \textbf{(f)}: $F(t)$ averaged over all times $t < 100\tau$. Comparing with the classical dynamics in \fref{figure:ClassicalQuantumDynamics} we see that quantum states in the classically chaotic region display a high degree of quantum information scrambling (i.e. growth of $C(t) = 1 - F(t)$), while quantum states in classically regular regions do not display the same amount of scrambling.}\label{figure:OTOCTimeEvolutionAlternate}
\end{figure*}

In \fref{figure:OTOCTimeEvolutionAlternate}(a-e) we display snapshots of $F(t)$ in a spin $J=41/2$ system for increasing evolution times $t$, with the perturbation angle $\epsilon = \pi/40$. Each point $(\theta,\phi)$ in the figures corresponds to an initial SCS $\ket{\theta,\phi}$. The OTOC is then obtained using \eqref{eq:UnravelingOTOC} and the procedure outlined in section~\ref{subsection:ChoosingW}. We observe that the OTOC $F(t)$ quickly decays into a low, steady state value for some initial states, whereas other initial states retain a high OTOC value throughout the time evolution. Figure~\ref{figure:OTOCTimeEvolutionAlternate}(f) displays an average over the OTOC values for $t < 100 \tau$. Together with the time evolution in \fref{figure:OTOCTimeEvolutionAlternate}(a-e), this reveals that the OTOC is minimal in a pattern corresponding to regions of many Floquet components (see \fref{figure:ClassicalQuantumDynamics}(b)), and that this pattern bears a remarkable resemblance to the classical chaotic region (see \fref{figure:ClassicalChaos}(a) and \fref{figure:ClassicalQuantumDynamics}(a)).

In \fref{figure:OTOCTimeEvolutionPlusSpinDependence}(b) we illustrate the time dependence of $F(t)$ for different initial SCS for $J=41/2$. The blue and red curves that originate in classically chaotic regions settle at oscillations around a reduced value, while the purple and yellow curves that correspond to classically regular regions retain their OTOC value close to unity throughout the time evolution. The mean values of the blue and red curves may be understood as the proportion of Hilbert space explored by these initial states during their time evolution, while the repeated revivals of the blue trajectory are due to the finite-sized Hilbert space, causing the partial return of the wave function to its initial state.

The green trajectory in \fref{figure:OTOCTimeEvolutionPlusSpinDependence}(b) is the OTOC $F(t)$ for the initial state $\ket{\pi/2,\pi/2}$ which -- due to the angular uncertainty of the SCS at $J=41/2$ -- partially overlaps with both regions of classically regular behavior and regions of classically chaotic behavior.
This trajectory displays oscillatory behavior far from that of both the blue and yellow trajectories, indicating that the regular behavior is suppressing chaos. The less-than-unity mean value of the green trajectory is due to the size of the Hilbert space on which the state spreads out during time evolution.

\subsection{Spin size dependence of single OTOC trajectories}\label{section:SingleOTOCTrajectories}
\begin{figure*}
\center
\includegraphics[width=0.85\linewidth]{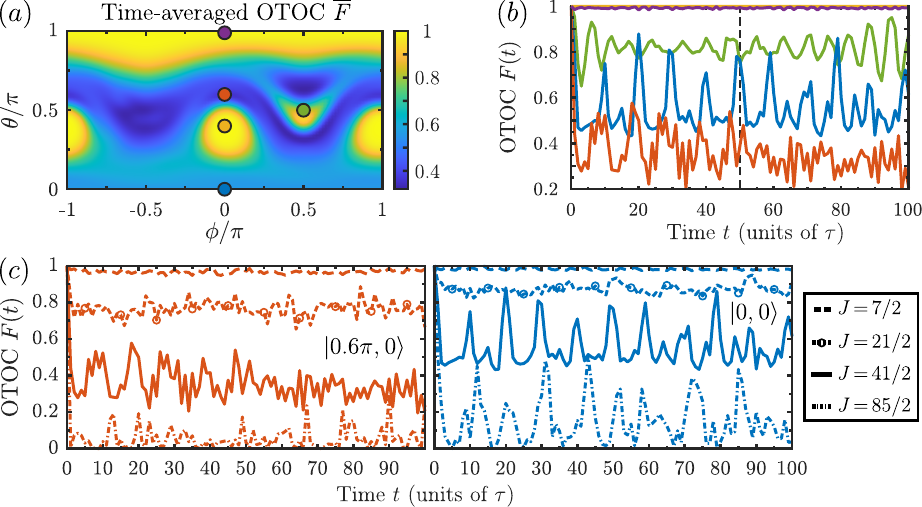}
\caption{OTOC $F(t)$ trajectories for different initial states and spin sizes. \textbf{(a):} OTOC $F(t)$ averaged over all times $t < 100\tau$, for spin $J=41/2$ and rotation angle $\epsilon = \pi/40$, identical to Fig.~\ref{figure:OTOCTimeEvolutionAlternate}(f). The colored circles identify the initial SCS locations in parameter space used in panels (b-c). \textbf{(b):} $F(t)$ trajectories for different initial SCS with $J=41/2$ and $\epsilon = \pi/40$. At $t = 50\tau$ the curves from top to bottom correspond to the following initial SCS: $\ket{0.4\pi,0}$ (yellow), $\ket{\pi,0}$ (purple), $\ket{\pi/2,\pi/2}$ (green), $\ket{0,0}$ (blue), and $\ket{0.6\pi,0}$ (red). The yellow and purple curves are classically regular and close to unity for all times, whereas the blue and red curves are classically chaotic. The initial state of the green trajectory overlaps significantly with both regular and chaotic regions, however the trajectory is dominated by the regular, oscillatory behavior. \textbf{(c):} $F(t)$ trajectories for varying spins $J$, for the chaotic initial SCS $\ket{0.6\pi,0}$ (left panel, red lines) and $\ket{0,0}$ (right panel, blue lines). In both panels the lines from top to bottom correspond to $J=7/2$ (dashed), $J=21/2$ (dashed with circles), $J=41/2$ (solid), and $J = 85/2$ (dash dotted). A clear spin dependence is visible in both panels, with lower spin sizes decreasing the visibility of the initial OTOC decay from unity.}
\label{figure:OTOCTimeEvolutionPlusSpinDependence}
\end{figure*}

The spin size $J$ influences the OTOC signal through the size of the Hilbert space ($\text{dim}\,\mathcal{H} = 2 J+1$) and the angular uncertainty of the SCS. The large uncertainty of small spins (e.g., the experimentally relevant $J = 7/2$~\cite{PhysRevE.98.042206,Nature.579.205}) could potentially destroy any visibility of quantum chaotic behavior. 
Here, we seek to investigate how the trajectory characteristics depend on the spin size~$J$.
In \fref{figure:OTOCTimeEvolutionPlusSpinDependence}(c) OTOC trajectories originating in the chaotic SCS $\ket{0.6\pi,0}$ (left) and $\ket{0,0}$ (right) are shown for several spin values. Both SCS show a dependence of their post-decay mean value on the spin size, which is expected as the Hilbert space dimension grows with $J$. In the right panel of \fref{figure:OTOCTimeEvolutionPlusSpinDependence}(c) we see that the periodic partial revivals do not survive into lower spin sizes.

\begin{figure}
\center
\includegraphics[width=0.85\linewidth]{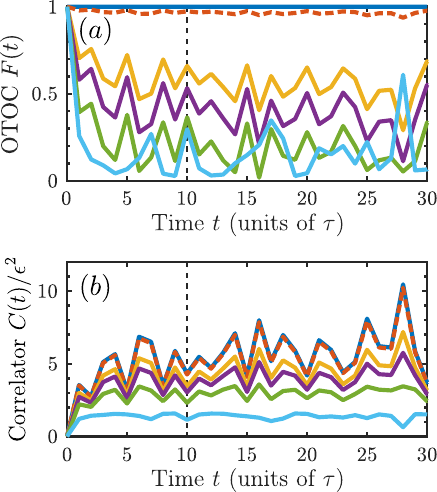}
\caption{$F(t)$ trajectories (a) and $C(t)/\epsilon^2$ trajectories (b) for different perturbation angles $\epsilon$, shown for the SCS $\ket{0.6\pi,0}$ (classically chaotic). We have used $J=7/2$. In both panels, the curves at the dashed vertical lines are from top to bottom: $\epsilon = \pi/400$ (solid dark blue), $\pi/40$ (dashed red), $\pi/10$ (solid yellow), $\pi/8$ (solid purple), $\pi/6$ (solid green), and $\pi/4$ (solid light blue).
A larger angle $\epsilon$ increases the visibility of the OTOC $F(t)$.
However, $W$ is no longer a small perturbation to the state for large $\epsilon$. (b) shows this deviation from the perturbation limit for increasing $\epsilon$.}\label{figure:OTOCTrajectorySeveralAngles}
\end{figure}

It is not obvious from \fref{figure:OTOCTimeEvolutionPlusSpinDependence}(c) that the case $J=7/2$ will yield appreciable signatures of quantum chaos due to the low relative visibility of the $F(t)$ signal when compared to higher spin sizes. However, the eigenvalues of our unitary operator $W_\epsilon(\Omega)$ depend on the angle of rotation $\epsilon$ which -- following the discussion in section~\ref{subsection:ChoosingW} -- is tunable post-experiment due to our OTOC measurement protocol not requiring the perturbation $W$ to be applied to the state at all. We can therefore increase the magnitude of variation of the OTOC $F(t)$ by changing $\epsilon$ post-experiment. This is illustrated in \fref{figure:OTOCTrajectorySeveralAngles} for the chaotic initial SCS $\ket{0.6\pi,0}$, where we have used the experimentally relevant $J = 7/2$~\cite{PhysRevE.98.042206}. When $\epsilon$ is increased from $\pi/400$ to $\pi/4$, the OTOC-signal modulation increases and resembles that of larger spins such as $J=41/2$ in \fref{figure:OTOCTimeEvolutionPlusSpinDependence}(b). However, the association with the Loschmidt echo protocol assumes a small perturbation, and one should be cautious when drawing conclusions from the larger $\epsilon$ (i.e. lower) curves in \fref{figure:OTOCTrajectorySeveralAngles}(a). By inserting $V = \rho_0$ and $W_\epsilon(\Omega)$ into \eqref{eq:ScramblingCommutator} and expanding $W_\epsilon$ to second order in $\epsilon \ll 1$, one may show that $C(t) \approx \epsilon^2 \sigma^2_{J^\prime}(t)$ \cite{NatCommun.10.1581}, where $\sigma^2_{J^\prime}(t) = \braket{J^{\prime\,2}(t)} - \braket{J^\prime(t)}^2$ is the variance of the angular momentum component $J^\prime$ along the $(\theta,\phi)$ direction. Figure~\ref{figure:OTOCTrajectorySeveralAngles}(b) illustrates $C(t)/\epsilon^2$ for the same choices of $\epsilon$ as in \fref{figure:OTOCTrajectorySeveralAngles}(a), and we find that for $\epsilon < \pi/6$ the general $C(t)$ dynamics are preserved quite well while still representing the perturbation limit of the Loschmidt echo protocol.

\section{Outlook}\label{section:Outlook}
In this article we have proposed a protocol for measuring out-of-time-ordered correlation functions (OTOCs) that does not require reversal of the time evolution. Unlike Loschmidt echo methods \cite{Garttner2017}, our protocol evaluates the OTOC as an operator expectation value in a forward time evolved state. The protocol may be applied either repeatedly on a single-system or on an ensemble of identical systems, and we have demonstrated an extension of the protocol that evaluates OTOCs in arbitrary mixed initial states. Additionally, we have illustrated how the customary restriction on the operator $V$ in the Loschmidt echo methods may be relaxed, allowing our protocol to measure the OTOC in \eqref{eq:OTOC} for arbitrary operators $V$ and $W$. Our protocol thus presents a versatile method of measuring OTOCs in experimental relevant settings.

An alternative protocol in Ref.~\cite{PhysRevX.9.021061} measures (Hermitian) observables $W(t)$ and $V^\dagger\, W(t)\, V$ from randomized initial states prepared using global random unitaries. The ensemble average of $\braket{W(t)}\braket{V^\dagger\, W(t)\, V}$ over random preparations then yields the OTOC $F(t)$ which, like our proposal, requires neither reversal of time evolution nor auxiliary degrees of freedom in the experimental protocol. However, the random unitary sampling restricts the evaluation to thermal states. In comparison, our proposed protocol can determine $F(t)$ for different pure initial states and, e.g., resolve regular and chaotic regions of state-space, while also being extendable to arbitrary mixed initial states.

We have illustrated our protocol by applying it in the analysis of quantum information scrambling within the quantum driven top. Our protocol readily finds application in the investigation of quantum information scrambling and quantum chaos in other experimentally relevant systems, such as by measuring the individual spins in ion trap quantum simulators \cite{nature.567.61,PhysRevLett.124.240505}, by measuring the single site occupation of atoms in optical lattices \cite{Nature.481.484}, or by measuring the expectation values of unitary operators in the hyperfine manifold of cold alkali atoms \cite{Nature.461.768}.

Finally, we recall that the protocol presented in this paper is restricted to the unitary time evolution of a closed quantum system. By comparing with the extended quantum regression theorem for out-of-time-ordered correlation functions in Ref.~\cite{PhysRevA.99.033816}, we note that the time evolution of \eqref{eq:UnravelingOTOC} omits deterministic contributions from environment observables inherent to OTOCs in open quantum systems, and the difficulty to reincorporate the irreversibly perturbed state of the environment and invert dissipation in the Loschmidt echo scheme complicates our approach in a similar manner. Still, we argue in section~\ref{subsection:OpenQuantumSystem} that our proposed protocol may be applied as a method of approximating the correct OTOC in open quantum systems. The validity of applying our protocol to open quantum systems is determined by the magnitude of the environment contributions and may be established on a case-by-case basis using the methods presented in Ref.~\cite{PhysRevA.99.033816}.

\section{Acknowledgments}
P.D.B. and K.M. acknowledge financial support from the Villum Foundation and from the Danish National
Research Foundation through the Center of Excellence
for Complex Quantum Systems (Grant agreement No.
DNRF156). S.A., V.M., M.A.I.J., and A.M. were funded by Australian Research Council Discovery Projects DP180100969 and DP210103769. V.M. acknowledges support from a Niels Stensen Fellowship. 

\appendix
\section{The Loschmidt echo as an out-of-time-ordered correlation function}
\label{App:LE-OTOC}
The Loschmidt echo provides an experimental protocol through which the quantum system dynamics' sensitivity to perturbations are probed. We let $\rho_0 = \ket{\Psi_0}\bra{\Psi_0}$ denote the pure initial system state at time $0$ and $W$ be a unitary operator applied at time $t$ to perturb the system. The resulting state of the Loschmidt echo is thus
\begin{equation}
    \rho_W(0;t) = W(t)\,\rho_0\,W^\dagger(t), \label{suppl:eq:LoschmidtEchoResultingState}
\end{equation}
where $W(t) = \mathcal{U}^\dagger(t,0)\,W\,\mathcal{U}(t,0)$ contains the Loschmidt echo procedure, as described in the main text. We want to compare $\rho_W(0;t)$ with the initial state $\rho_0$, hence the Loschmidt echo fidelity $L(t) = \text{Tr}[\rho_0\,\rho_W(0;t)]$ provides a suitable measure of the system dynamics' susceptibility to perturbation.

Let one operator $V(0) = \rho_0 = \ket{\Psi_0}\bra{\Psi_0}$ be the projection onto the pure initial state $\rho_0$. It then follows that
\begin{align}
    L(t) =& \text{Tr}[\rho_0\,\rho_W(0;t)] \nonumber\\
    =& \text{Tr}[\rho_0\,W(t)\,\rho_0\,W^\dagger(t)] \nonumber\\
    =& \text{Tr}[V(0)^\dagger\,W(t)\,V(0)\rho_0\,W^\dagger(t)] \nonumber\\ 
    =& \braket{W^\dagger(t)\,V(0)^\dagger\,W(t)\,V(0)} \nonumber\\
    =& F(t),
\end{align}
where we in the third equality have used that $V(0)^\dagger = V(0)$ and $V(0)\rho_0 = \rho_0$, and in the fourth equality have used the cyclic property of the trace. This shows us that the Loschmidt echo may provide direct experimental access to the OTOC $F(t)$, as demonstrated in Ref.~\cite{Garttner2017}.

\section{Estimating the Lyapunov exponent}\label{App:VariationalEquation}
The Lyapunov exponent $\lambda$ characterizes how an initial perturbation $
\delta x(0)$ at $t = 0$ evolves as $\delta x(t) \approx e^{\lambda t} \delta x(0)$.
A positive Lyapunov exponent $\lambda > 0$ indicates that a trajectory exponentially diverges from nearby states, which is a necessary condition for chaos.
By evolving an initial state and a variant with a slight perturbation $\delta x(0)$, the average Lyapunov exponent can be estimated as $\lambda \approx \log\left(\frac{|\delta x(t)|}{|\delta x(0)|}\right) t^{-1}$.
However, as the distance between a chaotic trajectory and its perturbed counterpart increases, it will at some point approach the size of the phase space, leading to finite-size effects.
Chaotic trajectories thus cannot be evolved indefinitely to obtain more and more accurate estimates of the Lyapunov exponent using this method.

An accurate estimate of the Lyapunov exponent that is insensitive to finite size effects can be determined by solving the variational equation~\cite{parker2012practical,chavez2016classical}
\begin{equation}
    \dot{\Phi}(x(0), t) = D_x f(x(t), t) \Phi(x(0), t),
\end{equation}
where $f(x, t)$ are the system equations, $D_x$ is the Jacobian matrix, i.e. the derivative with respect to coordinates.
The fundamental matrix $\Phi(x, t)$ dictates how an initial infinitesimal perturbation $\delta x(0)$ evolves in time via $\delta x(t) = \Phi(x(0), t) \delta x(0)$, and is subject to initial conditions $\Phi(t=0) = \mathbbm{1}$.

Since the variational equation depends on $x(t)$, a solution can be found by numerically integrating the variational equation alongside the equations of motion
\begin{equation}
    \begin{Bmatrix}
    \dot{x}\\
    \dot{\Phi}
    \end{Bmatrix} = 
    \begin{Bmatrix}
        f(x(t), t)\\
        D_x f(x(t), t) \Phi
    \end{Bmatrix},
\end{equation}
with initial conditions
\begin{equation}
    \begin{Bmatrix}
    x(0)\\
    \Phi(0)
    \end{Bmatrix} = 
    \begin{Bmatrix}
        x(0)\\
        \mathbbm{1}
    \end{Bmatrix}.
\end{equation}

The Lyapunov exponents displayed in \fref{figure:ClassicalChaos} were obtained by solving the variational equation for a total duration $T = 1000\tau$, where $\tau$ is the Floquet period.
The variational equation was solved for the range of initial coordinates $x(0) = [L_x, L_y, L_z]$ having spherical coordinates $(\phi, \theta)$.
The variational equation then relates an initial perturbation $\delta x(0)$ to the perturbation at a later point in time via $\delta x(T) = \Phi(T)\delta x(0)$.
A comparison of the initial and final perturbations provides an estimate of the Lyapunov exponent $\lambda \approx \log\left({\frac{\left|\Phi(T)\delta x(0)\right|}{\left|\delta x(0)\right|}}\right)T^{ -1}$.
For each $x(0)$ we chose 360 initial perturbations $\delta x(0)$ equally distributed along the plane orthogonal to $x(0)$ with magnitude $|\delta x(0)|=10^{-9}|x(0)|$, and averaged over the resulting Lyapunov exponents to obtain an average Lyapunov exponent.


%


\begin{thebibliography}{38}%
\makeatletter
\providecommand \@ifxundefined [1]{%
 \@ifx{#1\undefined}
}%
\providecommand \@ifnum [1]{%
 \ifnum #1\expandafter \@firstoftwo
 \else \expandafter \@secondoftwo
 \fi
}%
\providecommand \@ifx [1]{%
 \ifx #1\expandafter \@firstoftwo
 \else \expandafter \@secondoftwo
 \fi
}%
\providecommand \natexlab [1]{#1}%
\providecommand \enquote  [1]{``#1''}%
\providecommand \bibnamefont  [1]{#1}%
\providecommand \bibfnamefont [1]{#1}%
\providecommand \citenamefont [1]{#1}%
\providecommand \href@noop [0]{\@secondoftwo}%
\providecommand \href [0]{\begingroup \@sanitize@url \@href}%
\providecommand \@href[1]{\@@startlink{#1}\@@href}%
\providecommand \@@href[1]{\endgroup#1\@@endlink}%
\providecommand \@sanitize@url [0]{\catcode `\\12\catcode `\$12\catcode
  `\&12\catcode `\#12\catcode `\^12\catcode `\_12\catcode `\%12\relax}%
\providecommand \@@startlink[1]{}%
\providecommand \@@endlink[0]{}%
\providecommand \url  [0]{\begingroup\@sanitize@url \@url }%
\providecommand \@url [1]{\endgroup\@href {#1}{\urlprefix }}%
\providecommand \urlprefix  [0]{URL }%
\providecommand \Eprint [0]{\href }%
\providecommand \doibase [0]{http://dx.doi.org/}%
\providecommand \selectlanguage [0]{\@gobble}%
\providecommand \bibinfo  [0]{\@secondoftwo}%
\providecommand \bibfield  [0]{\@secondoftwo}%
\providecommand \translation [1]{[#1]}%
\providecommand \BibitemOpen [0]{}%
\providecommand \bibitemStop [0]{}%
\providecommand \bibitemNoStop [0]{.\EOS\space}%
\providecommand \EOS [0]{\spacefactor3000\relax}%
\providecommand \BibitemShut  [1]{\csname bibitem#1\endcsname}%
\let\auto@bib@innerbib\@empty
\bibitem [{\citenamefont {Lewis-Swan}\ \emph
  {et~al.}(2019{\natexlab{a}})\citenamefont {Lewis-Swan}, \citenamefont
  {Safavi-Naini}, \citenamefont {Bollinger},\ and\ \citenamefont
  {Rey}}]{NatCommun.10.1581}%
  \BibitemOpen
  \bibfield  {author} {\bibinfo {author} {\bibfnamefont {R.~J.}\ \bibnamefont
  {Lewis-Swan}}, \bibinfo {author} {\bibfnamefont {A.}~\bibnamefont
  {Safavi-Naini}}, \bibinfo {author} {\bibfnamefont {J.~J.}\ \bibnamefont
  {Bollinger}}, \ and\ \bibinfo {author} {\bibfnamefont {A.~M.}\ \bibnamefont
  {Rey}},\ }\bibfield  {title} {\enquote {\bibinfo {title} {Unifying
  scrambling, thermalization and entanglement through measurement of fidelity
  out-of-time-order correlators in the dicke model},}\ }\href {\doibase
  10.1038/s41467-019-09436-y} {\bibfield  {journal} {\bibinfo  {journal} {Nat.
  Commun.}\ }\textbf {\bibinfo {volume} {10}},\ \bibinfo {pages} {1581}
  (\bibinfo {year} {2019}{\natexlab{a}})}\BibitemShut {NoStop}%
\bibitem [{\citenamefont {Lewis-Swan}\ \emph
  {et~al.}(2019{\natexlab{b}})\citenamefont {Lewis-Swan}, \citenamefont
  {Safavi-Naini}, \citenamefont {Kaufman},\ and\ \citenamefont
  {Rey}}]{NatPhysRev.1.627}%
  \BibitemOpen
  \bibfield  {author} {\bibinfo {author} {\bibfnamefont {R.~J.}\ \bibnamefont
  {Lewis-Swan}}, \bibinfo {author} {\bibfnamefont {A.}~\bibnamefont
  {Safavi-Naini}}, \bibinfo {author} {\bibfnamefont {A.~M.}\ \bibnamefont
  {Kaufman}}, \ and\ \bibinfo {author} {\bibfnamefont {A.~M.}\ \bibnamefont
  {Rey}},\ }\bibfield  {title} {\enquote {\bibinfo {title} {Dynamics of quantum
  information},}\ }\href {\doibase 10.1038/s42254-019-0090-y} {\bibfield
  {journal} {\bibinfo  {journal} {Nat. Rev. Phys.}\ }\textbf {\bibinfo {volume}
  {1}},\ \bibinfo {pages} {627–634} (\bibinfo {year}
  {2019}{\natexlab{b}})}\BibitemShut {NoStop}%
\bibitem [{\citenamefont {Swingle}\ \emph {et~al.}(2016)\citenamefont
  {Swingle}, \citenamefont {Bentsen}, \citenamefont {Schleier-Smith},\ and\
  \citenamefont {Hayden}}]{PhysRevA.94.040302}%
  \BibitemOpen
  \bibfield  {author} {\bibinfo {author} {\bibfnamefont {B.}~\bibnamefont
  {Swingle}}, \bibinfo {author} {\bibfnamefont {G.}~\bibnamefont {Bentsen}},
  \bibinfo {author} {\bibfnamefont {M.}~\bibnamefont {Schleier-Smith}}, \ and\
  \bibinfo {author} {\bibfnamefont {P.}~\bibnamefont {Hayden}},\ }\bibfield
  {title} {\enquote {\bibinfo {title} {Measuring the scrambling of quantum
  information},}\ }\href {\doibase 10.1103/PhysRevA.94.040302} {\bibfield
  {journal} {\bibinfo  {journal} {Phys. Rev. A}\ }\textbf {\bibinfo {volume}
  {94}},\ \bibinfo {pages} {040302} (\bibinfo {year} {2016})}\BibitemShut
  {NoStop}%
\bibitem [{\citenamefont {Maldacena}\ \emph {et~al.}(2016)\citenamefont
  {Maldacena}, \citenamefont {Shenker},\ and\ \citenamefont
  {Stanford}}]{Maldacena2016}%
  \BibitemOpen
  \bibfield  {author} {\bibinfo {author} {\bibfnamefont {J.}~\bibnamefont
  {Maldacena}}, \bibinfo {author} {\bibfnamefont {S.~H.}\ \bibnamefont
  {Shenker}}, \ and\ \bibinfo {author} {\bibfnamefont {D.}~\bibnamefont
  {Stanford}},\ }\bibfield  {title} {\enquote {\bibinfo {title} {A bound on
  chaos},}\ }\href {\doibase 10.1007/JHEP08(2016)106} {\bibfield  {journal}
  {\bibinfo  {journal} {JHEP}\ }\textbf {\bibinfo {volume} {08}},\ \bibinfo
  {pages} {106} (\bibinfo {year} {2016})}\BibitemShut {NoStop}%
\bibitem [{\citenamefont {Fortes}\ \emph {et~al.}(2019)\citenamefont {Fortes},
  \citenamefont {Garc\'{\i}a-Mata}, \citenamefont {Jalabert},\ and\
  \citenamefont {Wisniacki}}]{PhysRevE.100.042201}%
  \BibitemOpen
  \bibfield  {author} {\bibinfo {author} {\bibfnamefont {E.~M.}\ \bibnamefont
  {Fortes}}, \bibinfo {author} {\bibfnamefont {I.}~\bibnamefont
  {Garc\'{\i}a-Mata}}, \bibinfo {author} {\bibfnamefont {R.~A.}\ \bibnamefont
  {Jalabert}}, \ and\ \bibinfo {author} {\bibfnamefont {D.~A.}\ \bibnamefont
  {Wisniacki}},\ }\bibfield  {title} {\enquote {\bibinfo {title} {Gauging
  classical and quantum integrability through out-of-time-ordered
  correlators},}\ }\href {\doibase 10.1103/PhysRevE.100.042201} {\bibfield
  {journal} {\bibinfo  {journal} {Phys. Rev. E}\ }\textbf {\bibinfo {volume}
  {100}},\ \bibinfo {pages} {042201} (\bibinfo {year} {2019})}\BibitemShut
  {NoStop}%
\bibitem [{\citenamefont {Larkin}\ and\ \citenamefont
  {Ovchinnikov}(1969)}]{Larkin1969}%
  \BibitemOpen
  \bibfield  {author} {\bibinfo {author} {\bibfnamefont {A.}~\bibnamefont
  {Larkin}}\ and\ \bibinfo {author} {\bibfnamefont {Y.~N.}\ \bibnamefont
  {Ovchinnikov}},\ }\bibfield  {title} {\enquote {\bibinfo {title}
  {Quasiclassical method in the theory of superconductivity},}\ }\href
  {http://www.jetp.ac.ru/cgi-bin/e/index/e/28/6/p1200?a=list} {\bibfield
  {journal} {\bibinfo  {journal} {JETP}\ }\textbf {\bibinfo {volume} {28}},\
  \bibinfo {pages} {1200} (\bibinfo {year} {1969})}\BibitemShut {NoStop}%
\bibitem [{\citenamefont {Marino}\ and\ \citenamefont
  {Rey}(2019)}]{PhysRevA.99.051803}%
  \BibitemOpen
  \bibfield  {author} {\bibinfo {author} {\bibfnamefont {J.}~\bibnamefont
  {Marino}}\ and\ \bibinfo {author} {\bibfnamefont {A.~M.}\ \bibnamefont
  {Rey}},\ }\bibfield  {title} {\enquote {\bibinfo {title} {Cavity-{QED}
  simulator of slow and fast scrambling},}\ }\href {\doibase
  10.1103/PhysRevA.99.051803} {\bibfield  {journal} {\bibinfo  {journal} {Phys.
  Rev. A}\ }\textbf {\bibinfo {volume} {99}},\ \bibinfo {pages} {051803}
  (\bibinfo {year} {2019})}\BibitemShut {NoStop}%
\bibitem [{\citenamefont {Xu}\ \emph {et~al.}(2020)\citenamefont {Xu},
  \citenamefont {Scaffidi},\ and\ \citenamefont
  {Cao}}]{PhysRevLett.124.140602}%
  \BibitemOpen
  \bibfield  {author} {\bibinfo {author} {\bibfnamefont {T.}~\bibnamefont
  {Xu}}, \bibinfo {author} {\bibfnamefont {T.}~\bibnamefont {Scaffidi}}, \ and\
  \bibinfo {author} {\bibfnamefont {X.}~\bibnamefont {Cao}},\ }\bibfield
  {title} {\enquote {\bibinfo {title} {Does scrambling equal chaos?}}\ }\href
  {\doibase 10.1103/PhysRevLett.124.140602} {\bibfield  {journal} {\bibinfo
  {journal} {Phys. Rev. Lett.}\ }\textbf {\bibinfo {volume} {124}},\ \bibinfo
  {pages} {140602} (\bibinfo {year} {2020})}\BibitemShut {NoStop}%
\bibitem [{\citenamefont {Kidd}\ \emph {et~al.}(2021)\citenamefont {Kidd},
  \citenamefont {Safavi-Naini},\ and\ \citenamefont
  {Corney}}]{PhysRevA.103.033304}%
  \BibitemOpen
  \bibfield  {author} {\bibinfo {author} {\bibfnamefont {R.~A.}\ \bibnamefont
  {Kidd}}, \bibinfo {author} {\bibfnamefont {A.}~\bibnamefont {Safavi-Naini}},
  \ and\ \bibinfo {author} {\bibfnamefont {J.~F.}\ \bibnamefont {Corney}},\
  }\bibfield  {title} {\enquote {\bibinfo {title} {Saddle-point scrambling
  without thermalization},}\ }\href {\doibase 10.1103/PhysRevA.103.033304}
  {\bibfield  {journal} {\bibinfo  {journal} {Phys. Rev. A}\ }\textbf {\bibinfo
  {volume} {103}},\ \bibinfo {pages} {033304} (\bibinfo {year}
  {2021})}\BibitemShut {NoStop}%
\bibitem [{\citenamefont {G\"arttner}\ \emph {et~al.}(2017)\citenamefont
  {G\"arttner}, \citenamefont {Bohnet}, \citenamefont {Safavi-Naini},
  \citenamefont {Wall}, \citenamefont {Bollinger},\ and\ \citenamefont
  {Rey}}]{Garttner2017}%
  \BibitemOpen
  \bibfield  {author} {\bibinfo {author} {\bibfnamefont {M.}~\bibnamefont
  {G\"arttner}}, \bibinfo {author} {\bibfnamefont {J.~G.}\ \bibnamefont
  {Bohnet}}, \bibinfo {author} {\bibfnamefont {A.}~\bibnamefont
  {Safavi-Naini}}, \bibinfo {author} {\bibfnamefont {M.~L.}\ \bibnamefont
  {Wall}}, \bibinfo {author} {\bibfnamefont {J.~J.}\ \bibnamefont {Bollinger}},
  \ and\ \bibinfo {author} {\bibfnamefont {A.~M.}\ \bibnamefont {Rey}},\
  }\bibfield  {title} {\enquote {\bibinfo {title} {Measuring out-of-time-order
  correlations and multiple quantum spectra in a trapped-ion quantum magnet},}\
  }\href {https://doi.org/10.1038/nphys4119} {\bibfield  {journal} {\bibinfo
  {journal} {Nature Physics}\ }\textbf {\bibinfo {volume} {13}},\ \bibinfo
  {pages} {781} (\bibinfo {year} {2017})}\BibitemShut {NoStop}%
\bibitem [{\citenamefont {G\"arttner}\ \emph {et~al.}(2018)\citenamefont
  {G\"arttner}, \citenamefont {Hauke},\ and\ \citenamefont
  {Rey}}]{PhysRevLett.120.040402}%
  \BibitemOpen
  \bibfield  {author} {\bibinfo {author} {\bibfnamefont {M.}~\bibnamefont
  {G\"arttner}}, \bibinfo {author} {\bibfnamefont {P.}~\bibnamefont {Hauke}}, \
  and\ \bibinfo {author} {\bibfnamefont {A.~M.}\ \bibnamefont {Rey}},\
  }\bibfield  {title} {\enquote {\bibinfo {title} {Relating out-of-time-order
  correlations to entanglement via multiple-quantum coherences},}\ }\href
  {\doibase 10.1103/PhysRevLett.120.040402} {\bibfield  {journal} {\bibinfo
  {journal} {Phys. Rev. Lett.}\ }\textbf {\bibinfo {volume} {120}},\ \bibinfo
  {pages} {040402} (\bibinfo {year} {2018})}\BibitemShut {NoStop}%
\bibitem [{\citenamefont {Li}\ \emph {et~al.}(2017)\citenamefont {Li},
  \citenamefont {Fan}, \citenamefont {Wang}, \citenamefont {Ye}, \citenamefont
  {Zeng}, \citenamefont {Zhai}, \citenamefont {Peng},\ and\ \citenamefont
  {Du}}]{PhysRevX.7.031011}%
  \BibitemOpen
  \bibfield  {author} {\bibinfo {author} {\bibfnamefont {J.}~\bibnamefont
  {Li}}, \bibinfo {author} {\bibfnamefont {R.}~\bibnamefont {Fan}}, \bibinfo
  {author} {\bibfnamefont {H.}~\bibnamefont {Wang}}, \bibinfo {author}
  {\bibfnamefont {B.}~\bibnamefont {Ye}}, \bibinfo {author} {\bibfnamefont
  {B.}~\bibnamefont {Zeng}}, \bibinfo {author} {\bibfnamefont {H.}~\bibnamefont
  {Zhai}}, \bibinfo {author} {\bibfnamefont {X.}~\bibnamefont {Peng}}, \ and\
  \bibinfo {author} {\bibfnamefont {J.}~\bibnamefont {Du}},\ }\bibfield
  {title} {\enquote {\bibinfo {title} {Measuring out-of-time-order correlators
  on a nuclear magnetic resonance quantum simulator},}\ }\href {\doibase
  10.1103/PhysRevX.7.031011} {\bibfield  {journal} {\bibinfo  {journal} {Phys.
  Rev. X}\ }\textbf {\bibinfo {volume} {7}},\ \bibinfo {pages} {031011}
  (\bibinfo {year} {2017})}\BibitemShut {NoStop}%
\bibitem [{\citenamefont {Wei}\ \emph {et~al.}(2019)\citenamefont {Wei},
  \citenamefont {Peng}, \citenamefont {Shtanko}, \citenamefont {Marvian},
  \citenamefont {Lloyd}, \citenamefont {Ramanathan},\ and\ \citenamefont
  {Cappellaro}}]{PhysRevLett.123.090605}%
  \BibitemOpen
  \bibfield  {author} {\bibinfo {author} {\bibfnamefont {K.~X.}\ \bibnamefont
  {Wei}}, \bibinfo {author} {\bibfnamefont {P.}~\bibnamefont {Peng}}, \bibinfo
  {author} {\bibfnamefont {O.}~\bibnamefont {Shtanko}}, \bibinfo {author}
  {\bibfnamefont {I.}~\bibnamefont {Marvian}}, \bibinfo {author} {\bibfnamefont
  {S.}~\bibnamefont {Lloyd}}, \bibinfo {author} {\bibfnamefont
  {C.}~\bibnamefont {Ramanathan}}, \ and\ \bibinfo {author} {\bibfnamefont
  {P.}~\bibnamefont {Cappellaro}},\ }\bibfield  {title} {\enquote {\bibinfo
  {title} {Emergent prethermalization signatures in out-of-time ordered
  correlations},}\ }\href {\doibase 10.1103/PhysRevLett.123.090605} {\bibfield
  {journal} {\bibinfo  {journal} {Phys. Rev. Lett.}\ }\textbf {\bibinfo
  {volume} {123}},\ \bibinfo {pages} {090605} (\bibinfo {year}
  {2019})}\BibitemShut {NoStop}%
\bibitem [{\citenamefont {Yao}\ \emph {et~al.}(2016)\citenamefont {Yao},
  \citenamefont {Grusdt}, \citenamefont {Swingle}, \citenamefont {Lukin},
  \citenamefont {Stamper-Kurn}, \citenamefont {Moore},\ and\ \citenamefont
  {Demler}}]{arXiv.1607.01801}%
  \BibitemOpen
  \bibfield  {author} {\bibinfo {author} {\bibfnamefont {N.~Y.}\ \bibnamefont
  {Yao}}, \bibinfo {author} {\bibfnamefont {F.}~\bibnamefont {Grusdt}},
  \bibinfo {author} {\bibfnamefont {B.}~\bibnamefont {Swingle}}, \bibinfo
  {author} {\bibfnamefont {M.~D.}\ \bibnamefont {Lukin}}, \bibinfo {author}
  {\bibfnamefont {D.~M.}\ \bibnamefont {Stamper-Kurn}}, \bibinfo {author}
  {\bibfnamefont {J.~E.}\ \bibnamefont {Moore}}, \ and\ \bibinfo {author}
  {\bibfnamefont {E.~A.}\ \bibnamefont {Demler}},\ }\bibfield  {title}
  {\enquote {\bibinfo {title} {Interferometric approach to probing fast
  scrambling},}\ }\href {https://arxiv.org/abs/1607.01801} {\bibfield
  {journal} {\bibinfo  {journal} {arXiv:1607.01801}\ } (\bibinfo {year}
  {2016})}\BibitemShut {NoStop}%
\bibitem [{\citenamefont {Landsman}\ \emph {et~al.}(2019)\citenamefont
  {Landsman}, \citenamefont {Figgatt}, \citenamefont {Schuster}, \citenamefont
  {Linke}, \citenamefont {Yoshida}, \citenamefont {Yao},\ and\ \citenamefont
  {Monroe}}]{nature.567.61}%
  \BibitemOpen
  \bibfield  {author} {\bibinfo {author} {\bibfnamefont {K.~A.}\ \bibnamefont
  {Landsman}}, \bibinfo {author} {\bibfnamefont {C.}~\bibnamefont {Figgatt}},
  \bibinfo {author} {\bibfnamefont {T.}~\bibnamefont {Schuster}}, \bibinfo
  {author} {\bibfnamefont {N.~M.}\ \bibnamefont {Linke}}, \bibinfo {author}
  {\bibfnamefont {B.}~\bibnamefont {Yoshida}}, \bibinfo {author} {\bibfnamefont
  {N.~Y.}\ \bibnamefont {Yao}}, \ and\ \bibinfo {author} {\bibfnamefont
  {C.}~\bibnamefont {Monroe}},\ }\bibfield  {title} {\enquote {\bibinfo {title}
  {Verified quantum information scrambling},}\ }\href {\doibase
  10.1038/s41586-019-0952-6} {\bibfield  {journal} {\bibinfo  {journal}
  {Nature}\ }\textbf {\bibinfo {volume} {567}},\ \bibinfo {pages} {61--65}
  (\bibinfo {year} {2019})}\BibitemShut {NoStop}%
\bibitem [{\citenamefont {Vermersch}\ \emph {et~al.}(2019)\citenamefont
  {Vermersch}, \citenamefont {Elben}, \citenamefont {Sieberer}, \citenamefont
  {Yao},\ and\ \citenamefont {Zoller}}]{PhysRevX.9.021061}%
  \BibitemOpen
  \bibfield  {author} {\bibinfo {author} {\bibfnamefont {B.}~\bibnamefont
  {Vermersch}}, \bibinfo {author} {\bibfnamefont {A.}~\bibnamefont {Elben}},
  \bibinfo {author} {\bibfnamefont {L.~M.}\ \bibnamefont {Sieberer}}, \bibinfo
  {author} {\bibfnamefont {N.~Y.}\ \bibnamefont {Yao}}, \ and\ \bibinfo
  {author} {\bibfnamefont {P.}~\bibnamefont {Zoller}},\ }\bibfield  {title}
  {\enquote {\bibinfo {title} {Probing scrambling using statistical
  correlations between randomized measurements},}\ }\href {\doibase
  10.1103/PhysRevX.9.021061} {\bibfield  {journal} {\bibinfo  {journal} {Phys.
  Rev. X}\ }\textbf {\bibinfo {volume} {9}},\ \bibinfo {pages} {021061}
  (\bibinfo {year} {2019})}\BibitemShut {NoStop}%
\bibitem [{\citenamefont {Joshi}\ \emph {et~al.}(2020)\citenamefont {Joshi},
  \citenamefont {Elben}, \citenamefont {Vermersch}, \citenamefont {Brydges},
  \citenamefont {Maier}, \citenamefont {Zoller}, \citenamefont {Blatt},\ and\
  \citenamefont {Roos}}]{PhysRevLett.124.240505}%
  \BibitemOpen
  \bibfield  {author} {\bibinfo {author} {\bibfnamefont {M.~K.}\ \bibnamefont
  {Joshi}}, \bibinfo {author} {\bibfnamefont {A.}~\bibnamefont {Elben}},
  \bibinfo {author} {\bibfnamefont {B.}~\bibnamefont {Vermersch}}, \bibinfo
  {author} {\bibfnamefont {T.}~\bibnamefont {Brydges}}, \bibinfo {author}
  {\bibfnamefont {C.}~\bibnamefont {Maier}}, \bibinfo {author} {\bibfnamefont
  {P.}~\bibnamefont {Zoller}}, \bibinfo {author} {\bibfnamefont
  {R.}~\bibnamefont {Blatt}}, \ and\ \bibinfo {author} {\bibfnamefont {C.~F.}\
  \bibnamefont {Roos}},\ }\bibfield  {title} {\enquote {\bibinfo {title}
  {Quantum information scrambling in a trapped-ion quantum simulator with
  tunable range interactions},}\ }\href {\doibase
  10.1103/PhysRevLett.124.240505} {\bibfield  {journal} {\bibinfo  {journal}
  {Phys. Rev. Lett.}\ }\textbf {\bibinfo {volume} {124}},\ \bibinfo {pages}
  {240505} (\bibinfo {year} {2020})}\BibitemShut {NoStop}%
\bibitem [{\citenamefont {Mourik}\ \emph {et~al.}(2018)\citenamefont {Mourik},
  \citenamefont {Asaad}, \citenamefont {Firgau}, \citenamefont {Pla},
  \citenamefont {Holmes}, \citenamefont {Milburn}, \citenamefont {McCallum},\
  and\ \citenamefont {Morello}}]{PhysRevE.98.042206}%
  \BibitemOpen
  \bibfield  {author} {\bibinfo {author} {\bibfnamefont {V.}~\bibnamefont
  {Mourik}}, \bibinfo {author} {\bibfnamefont {S.}~\bibnamefont {Asaad}},
  \bibinfo {author} {\bibfnamefont {H.}~\bibnamefont {Firgau}}, \bibinfo
  {author} {\bibfnamefont {J.~J.}\ \bibnamefont {Pla}}, \bibinfo {author}
  {\bibfnamefont {C.}~\bibnamefont {Holmes}}, \bibinfo {author} {\bibfnamefont
  {G.~J.}\ \bibnamefont {Milburn}}, \bibinfo {author} {\bibfnamefont {J.~C.}\
  \bibnamefont {McCallum}}, \ and\ \bibinfo {author} {\bibfnamefont
  {A.}~\bibnamefont {Morello}},\ }\bibfield  {title} {\enquote {\bibinfo
  {title} {Exploring quantum chaos with a single nuclear spin},}\ }\href
  {\doibase 10.1103/PhysRevE.98.042206} {\bibfield  {journal} {\bibinfo
  {journal} {Phys. Rev. E}\ }\textbf {\bibinfo {volume} {98}},\ \bibinfo
  {pages} {042206} (\bibinfo {year} {2018})}\BibitemShut {NoStop}%
\bibitem [{\citenamefont {Gorin}\ \emph {et~al.}(2006)\citenamefont {Gorin},
  \citenamefont {Prosen}, \citenamefont {Seligman},\ and\ \citenamefont
  {Žnidarič}}]{GORIN200633}%
  \BibitemOpen
  \bibfield  {author} {\bibinfo {author} {\bibfnamefont {T.}~\bibnamefont
  {Gorin}}, \bibinfo {author} {\bibfnamefont {T.}~\bibnamefont {Prosen}},
  \bibinfo {author} {\bibfnamefont {T.~H.}\ \bibnamefont {Seligman}}, \ and\
  \bibinfo {author} {\bibfnamefont {M.}~\bibnamefont {Žnidarič}},\ }\bibfield
   {title} {\enquote {\bibinfo {title} {Dynamics of {L}oschmidt echoes and
  fidelity decay},}\ }\href {\doibase
  https://doi.org/10.1016/j.physrep.2006.09.003} {\bibfield  {journal}
  {\bibinfo  {journal} {Physics Reports}\ }\textbf {\bibinfo {volume} {435}},\
  \bibinfo {pages} {33} (\bibinfo {year} {2006})}\BibitemShut {NoStop}%
\bibitem [{\citenamefont {Blocher}\ and\ \citenamefont
  {M\o{}lmer}(2019)}]{PhysRevA.99.033816}%
  \BibitemOpen
  \bibfield  {author} {\bibinfo {author} {\bibfnamefont {P.~D.}\ \bibnamefont
  {Blocher}}\ and\ \bibinfo {author} {\bibfnamefont {K.}~\bibnamefont
  {M\o{}lmer}},\ }\bibfield  {title} {\enquote {\bibinfo {title} {Quantum
  regression theorem for out-of-time-ordered correlation functions},}\ }\href
  {\doibase 10.1103/PhysRevA.99.033816} {\bibfield  {journal} {\bibinfo
  {journal} {Phys. Rev. A}\ }\textbf {\bibinfo {volume} {99}},\ \bibinfo
  {pages} {033816} (\bibinfo {year} {2019})}\BibitemShut {NoStop}%
\bibitem [{\citenamefont {Harris}\ \emph {et~al.}(2022)\citenamefont {Harris},
  \citenamefont {Yan},\ and\ \citenamefont
  {Sinitsyn}}]{PhysRevLett.129.050602}%
  \BibitemOpen
  \bibfield  {author} {\bibinfo {author} {\bibfnamefont {Joseph}\ \bibnamefont
  {Harris}}, \bibinfo {author} {\bibfnamefont {Bin}\ \bibnamefont {Yan}}, \
  and\ \bibinfo {author} {\bibfnamefont {Nikolai~A.}\ \bibnamefont
  {Sinitsyn}},\ }\bibfield  {title} {\enquote {\bibinfo {title} {Benchmarking
  information scrambling},}\ }\href {\doibase 10.1103/PhysRevLett.129.050602}
  {\bibfield  {journal} {\bibinfo  {journal} {Phys. Rev. Lett.}\ }\textbf
  {\bibinfo {volume} {129}},\ \bibinfo {pages} {050602} (\bibinfo {year}
  {2022})}\BibitemShut {NoStop}%
\bibitem [{\citenamefont {Haake}\ \emph {et~al.}(1990)\citenamefont {Haake},
  \citenamefont {Lenz},\ and\ \citenamefont {Puri}}]{haake1990optical}%
  \BibitemOpen
  \bibfield  {author} {\bibinfo {author} {\bibfnamefont {F.}~\bibnamefont
  {Haake}}, \bibinfo {author} {\bibfnamefont {G.}~\bibnamefont {Lenz}}, \ and\
  \bibinfo {author} {\bibfnamefont {R.}~\bibnamefont {Puri}},\ }\bibfield
  {title} {\enquote {\bibinfo {title} {Optical tops},}\ }\href {\doibase
  https://doi.org/10.1080/09500349014550221} {\bibfield  {journal} {\bibinfo
  {journal} {Journal of Modern Optics}\ }\textbf {\bibinfo {volume} {37}},\
  \bibinfo {pages} {155--158} (\bibinfo {year} {1990})}\BibitemShut {NoStop}%
\bibitem [{\citenamefont {Haake}\ \emph {et~al.}(1987)\citenamefont {Haake},
  \citenamefont {Ku{\'s}},\ and\ \citenamefont {Scharf}}]{haake1987classical}%
  \BibitemOpen
  \bibfield  {author} {\bibinfo {author} {\bibfnamefont {F.}~\bibnamefont
  {Haake}}, \bibinfo {author} {\bibfnamefont {M.}~\bibnamefont {Ku{\'s}}}, \
  and\ \bibinfo {author} {\bibfnamefont {R.}~\bibnamefont {Scharf}},\
  }\bibfield  {title} {\enquote {\bibinfo {title} {Classical and quantum chaos
  for a kicked top},}\ }\href {\doibase https://doi.org/10.1007/BF01303727}
  {\bibfield  {journal} {\bibinfo  {journal} {Z. Physik B - Condensed Matter}\
  }\textbf {\bibinfo {volume} {65}},\ \bibinfo {pages} {381--395} (\bibinfo
  {year} {1987})}\BibitemShut {NoStop}%
\bibitem [{\citenamefont {Scharf}\ \emph {et~al.}(1988)\citenamefont {Scharf},
  \citenamefont {Dietz}, \citenamefont {Ku{\'s}}, \citenamefont {Haake},\ and\
  \citenamefont {Berry}}]{scharf1988kramers}%
  \BibitemOpen
  \bibfield  {author} {\bibinfo {author} {\bibfnamefont {R.}~\bibnamefont
  {Scharf}}, \bibinfo {author} {\bibfnamefont {B.}~\bibnamefont {Dietz}},
  \bibinfo {author} {\bibfnamefont {M.}~\bibnamefont {Ku{\'s}}}, \bibinfo
  {author} {\bibfnamefont {F.}~\bibnamefont {Haake}}, \ and\ \bibinfo {author}
  {\bibfnamefont {M.~V.}\ \bibnamefont {Berry}},\ }\bibfield  {title} {\enquote
  {\bibinfo {title} {Kramers' degeneracy and quartic level repulsion},}\ }\href
  {\doibase https://doi.org/10.1209/0295-5075/5/5/001} {\bibfield  {journal}
  {\bibinfo  {journal} {EPL (Europhysics Letters)}\ }\textbf {\bibinfo {volume}
  {5}},\ \bibinfo {pages} {383} (\bibinfo {year} {1988})}\BibitemShut {NoStop}%
\bibitem [{\citenamefont {Zyczkowski}(1990)}]{zyczkowski1990indicators}%
  \BibitemOpen
  \bibfield  {author} {\bibinfo {author} {\bibfnamefont {K.}~\bibnamefont
  {Zyczkowski}},\ }\bibfield  {title} {\enquote {\bibinfo {title} {Indicators
  of quantum chaos based on eigenvector statistics},}\ }\href {\doibase
  https://doi.org/10.1088/0305-4470/23/20/005} {\bibfield  {journal} {\bibinfo
  {journal} {J. Phys. A: Math. Gen.}\ }\textbf {\bibinfo {volume} {23}},\
  \bibinfo {pages} {4427} (\bibinfo {year} {1990})}\BibitemShut {NoStop}%
\bibitem [{\citenamefont {Ku{\'s}}\ \emph {et~al.}(1991)\citenamefont
  {Ku{\'s}}, \citenamefont {Zakrzewski},\ and\ \citenamefont
  {{\.Z}yczkowski}}]{kus1991quantum}%
  \BibitemOpen
  \bibfield  {author} {\bibinfo {author} {\bibfnamefont {M.}~\bibnamefont
  {Ku{\'s}}}, \bibinfo {author} {\bibfnamefont {J.}~\bibnamefont {Zakrzewski}},
  \ and\ \bibinfo {author} {\bibfnamefont {K.}~\bibnamefont {{\.Z}yczkowski}},\
  }\bibfield  {title} {\enquote {\bibinfo {title} {Quantum scars on a
  sphere},}\ }\href {\doibase https://doi.org/10.1103/PhysRevA.43.4244}
  {\bibfield  {journal} {\bibinfo  {journal} {Phys. Rev. A}\ }\textbf {\bibinfo
  {volume} {43}},\ \bibinfo {pages} {4244} (\bibinfo {year}
  {1991})}\BibitemShut {NoStop}%
\bibitem [{\citenamefont {Schack}\ \emph {et~al.}(1994)\citenamefont {Schack},
  \citenamefont {D’Ariano},\ and\ \citenamefont
  {Caves}}]{schack1994hypersensitivity}%
  \BibitemOpen
  \bibfield  {author} {\bibinfo {author} {\bibfnamefont {R.}~\bibnamefont
  {Schack}}, \bibinfo {author} {\bibfnamefont {G.~M.}\ \bibnamefont
  {D’Ariano}}, \ and\ \bibinfo {author} {\bibfnamefont {C.~M.}\ \bibnamefont
  {Caves}},\ }\bibfield  {title} {\enquote {\bibinfo {title} {Hypersensitivity
  to perturbation in the quantum kicked top},}\ }\href {\doibase
  https://doi.org/10.1103/PhysRevE.50.972} {\bibfield  {journal} {\bibinfo
  {journal} {Phys. Rev. E}\ }\textbf {\bibinfo {volume} {50}},\ \bibinfo
  {pages} {972} (\bibinfo {year} {1994})}\BibitemShut {NoStop}%
\bibitem [{\citenamefont {Jacquod}\ \emph {et~al.}(2001)\citenamefont
  {Jacquod}, \citenamefont {Silvestrov},\ and\ \citenamefont
  {Beenakker}}]{jacquod2001golden}%
  \BibitemOpen
  \bibfield  {author} {\bibinfo {author} {\bibfnamefont {Ph.}\ \bibnamefont
  {Jacquod}}, \bibinfo {author} {\bibfnamefont {P.~G.}\ \bibnamefont
  {Silvestrov}}, \ and\ \bibinfo {author} {\bibfnamefont {C.~W.~J.}\
  \bibnamefont {Beenakker}},\ }\bibfield  {title} {\enquote {\bibinfo {title}
  {Golden rule decay versus lyapunov decay of the quantum loschmidt echo},}\
  }\href {\doibase https://doi.org/10.1103/PhysRevE.64.055203} {\bibfield
  {journal} {\bibinfo  {journal} {Phys. Rev. E}\ }\textbf {\bibinfo {volume}
  {64}},\ \bibinfo {pages} {055203} (\bibinfo {year} {2001})}\BibitemShut
  {NoStop}%
\bibitem [{\citenamefont {Holthaus}(2015)}]{JPhysB.49.013001}%
  \BibitemOpen
  \bibfield  {author} {\bibinfo {author} {\bibfnamefont {M.}~\bibnamefont
  {Holthaus}},\ }\bibfield  {title} {\enquote {\bibinfo {title} {Floquet
  engineering with quasienergy bands of periodically driven optical
  lattices},}\ }\href {\doibase https://doi.org/10.1088/0953-4075/49/1/013001}
  {\bibfield  {journal} {\bibinfo  {journal} {J. Phys. B: At. Mol. Opt. Phys.}\
  }\textbf {\bibinfo {volume} {49}},\ \bibinfo {pages} {013001} (\bibinfo
  {year} {2015})}\BibitemShut {NoStop}%
\bibitem [{\citenamefont {Yan}\ \emph {et~al.}(2020)\citenamefont {Yan},
  \citenamefont {Cincio},\ and\ \citenamefont
  {Zurek}}]{PhysRevLett.124.160603}%
  \BibitemOpen
  \bibfield  {author} {\bibinfo {author} {\bibfnamefont {B.}~\bibnamefont
  {Yan}}, \bibinfo {author} {\bibfnamefont {L.}~\bibnamefont {Cincio}}, \ and\
  \bibinfo {author} {\bibfnamefont {W.~H.}\ \bibnamefont {Zurek}},\ }\bibfield
  {title} {\enquote {\bibinfo {title} {Information scrambling and loschmidt
  echo},}\ }\href {\doibase 10.1103/PhysRevLett.124.160603} {\bibfield
  {journal} {\bibinfo  {journal} {Phys. Rev. Lett.}\ }\textbf {\bibinfo
  {volume} {124}},\ \bibinfo {pages} {160603} (\bibinfo {year}
  {2020})}\BibitemShut {NoStop}%
\bibitem [{\citenamefont {Arecchi}\ \emph {et~al.}(1972)\citenamefont
  {Arecchi}, \citenamefont {Courtens}, \citenamefont {Gilmore},\ and\
  \citenamefont {Thomas}}]{PhysRevA.6.2211}%
  \BibitemOpen
  \bibfield  {author} {\bibinfo {author} {\bibfnamefont {F.~T.}\ \bibnamefont
  {Arecchi}}, \bibinfo {author} {\bibfnamefont {E.}~\bibnamefont {Courtens}},
  \bibinfo {author} {\bibfnamefont {R.}~\bibnamefont {Gilmore}}, \ and\
  \bibinfo {author} {\bibfnamefont {H.}~\bibnamefont {Thomas}},\ }\bibfield
  {title} {\enquote {\bibinfo {title} {Atomic coherent states in quantum
  optics},}\ }\href {\doibase 10.1103/PhysRevA.6.2211} {\bibfield  {journal}
  {\bibinfo  {journal} {Phys. Rev. A}\ }\textbf {\bibinfo {volume} {6}},\
  \bibinfo {pages} {2211--2237} (\bibinfo {year} {1972})}\BibitemShut {NoStop}%
\bibitem [{\citenamefont {Sakurai}\ and\ \citenamefont
  {Napolitano}(2011)}]{Sakurai2011}%
  \BibitemOpen
  \bibfield  {author} {\bibinfo {author} {\bibfnamefont {J.~J.}\ \bibnamefont
  {Sakurai}}\ and\ \bibinfo {author} {\bibfnamefont {J.}~\bibnamefont
  {Napolitano}},\ }\href@noop {} {\emph {\bibinfo {title} {Modern quantum
  mechanics}}},\ \bibinfo {edition} {2nd}\ ed.\ (\bibinfo  {publisher} {San
  Francisco: Addison-Wesley},\ \bibinfo {year} {2011})\BibitemShut {NoStop}%
\bibitem [{\citenamefont {Sieberer}\ \emph {et~al.}(2019)\citenamefont
  {Sieberer}, \citenamefont {Olsacher}, \citenamefont {Elben}, \citenamefont
  {Heyl}, \citenamefont {Hauke}, \citenamefont {Haake},\ and\ \citenamefont
  {Zoller}}]{npjQuantInf.5.78}%
  \BibitemOpen
  \bibfield  {author} {\bibinfo {author} {\bibfnamefont {L.~M.}\ \bibnamefont
  {Sieberer}}, \bibinfo {author} {\bibfnamefont {T.}~\bibnamefont {Olsacher}},
  \bibinfo {author} {\bibfnamefont {A.}~\bibnamefont {Elben}}, \bibinfo
  {author} {\bibfnamefont {M.}~\bibnamefont {Heyl}}, \bibinfo {author}
  {\bibfnamefont {P.}~\bibnamefont {Hauke}}, \bibinfo {author} {\bibfnamefont
  {F.}~\bibnamefont {Haake}}, \ and\ \bibinfo {author} {\bibfnamefont
  {P.}~\bibnamefont {Zoller}},\ }\bibfield  {title} {\enquote {\bibinfo {title}
  {Digital quantum simulation, {T}rotter errors, and quantum chaos of the
  kicked top},}\ }\href {\doibase 10.1038/s41534-019-0192-5} {\bibfield
  {journal} {\bibinfo  {journal} {npj Quantum Inf}\ }\textbf {\bibinfo {volume}
  {5}},\ \bibinfo {pages} {78} (\bibinfo {year} {2019})}\BibitemShut {NoStop}%
\bibitem [{\citenamefont {Asaad}\ \emph {et~al.}(2020)\citenamefont {Asaad},
  \citenamefont {Mourik}, \citenamefont {Joecker}, \citenamefont {Johnson},
  \citenamefont {Baczewski}, \citenamefont {Firgau}, \citenamefont {Mądzik},
  \citenamefont {Schmitt}, \citenamefont {Pla}, \citenamefont {Hudson},
  \citenamefont {Itoh}, \citenamefont {McCallum}, \citenamefont {Dzurak},
  \citenamefont {Laucht},\ and\ \citenamefont {Morello}}]{Nature.579.205}%
  \BibitemOpen
  \bibfield  {author} {\bibinfo {author} {\bibfnamefont {S.}~\bibnamefont
  {Asaad}}, \bibinfo {author} {\bibfnamefont {V.}~\bibnamefont {Mourik}},
  \bibinfo {author} {\bibfnamefont {B.}~\bibnamefont {Joecker}}, \bibinfo
  {author} {\bibfnamefont {M.~A.~I.}\ \bibnamefont {Johnson}}, \bibinfo
  {author} {\bibfnamefont {A.~D.}\ \bibnamefont {Baczewski}}, \bibinfo {author}
  {\bibfnamefont {H.~R.}\ \bibnamefont {Firgau}}, \bibinfo {author}
  {\bibfnamefont {M.~T.}\ \bibnamefont {Mądzik}}, \bibinfo {author}
  {\bibfnamefont {V.}~\bibnamefont {Schmitt}}, \bibinfo {author} {\bibfnamefont
  {J.~J.}\ \bibnamefont {Pla}}, \bibinfo {author} {\bibfnamefont {F.~E.}\
  \bibnamefont {Hudson}}, \bibinfo {author} {\bibfnamefont {K.~M.}\
  \bibnamefont {Itoh}}, \bibinfo {author} {\bibfnamefont {J.~C.}\ \bibnamefont
  {McCallum}}, \bibinfo {author} {\bibfnamefont {A.~S.}\ \bibnamefont
  {Dzurak}}, \bibinfo {author} {\bibfnamefont {A.}~\bibnamefont {Laucht}}, \
  and\ \bibinfo {author} {\bibfnamefont {A.}~\bibnamefont {Morello}},\
  }\bibfield  {title} {\enquote {\bibinfo {title} {Coherent electrical control
  of a single high-spin nucleus in silicon},}\ }\href {\doibase
  10.1038/s41586-020-2057-7} {\bibfield  {journal} {\bibinfo  {journal}
  {Nature}\ }\textbf {\bibinfo {volume} {579}},\ \bibinfo {pages} {205--209}
  (\bibinfo {year} {2020})}\BibitemShut {NoStop}%
\bibitem [{\citenamefont {Cheneau}\ \emph {et~al.}(2012)\citenamefont
  {Cheneau}, \citenamefont {Barmettler}, \citenamefont {Poletti}, \citenamefont
  {Endres}, \citenamefont {Schauß}, \citenamefont {Fukuhara}, \citenamefont
  {Gross}, \citenamefont {Bloch}, \citenamefont {Kollath},\ and\ \citenamefont
  {Kuhr}}]{Nature.481.484}%
  \BibitemOpen
  \bibfield  {author} {\bibinfo {author} {\bibfnamefont {M.}~\bibnamefont
  {Cheneau}}, \bibinfo {author} {\bibfnamefont {P.}~\bibnamefont {Barmettler}},
  \bibinfo {author} {\bibfnamefont {D.}~\bibnamefont {Poletti}}, \bibinfo
  {author} {\bibfnamefont {M.}~\bibnamefont {Endres}}, \bibinfo {author}
  {\bibfnamefont {P.}~\bibnamefont {Schauß}}, \bibinfo {author} {\bibfnamefont
  {T.}~\bibnamefont {Fukuhara}}, \bibinfo {author} {\bibfnamefont
  {C.}~\bibnamefont {Gross}}, \bibinfo {author} {\bibfnamefont
  {I.}~\bibnamefont {Bloch}}, \bibinfo {author} {\bibfnamefont
  {C.}~\bibnamefont {Kollath}}, \ and\ \bibinfo {author} {\bibfnamefont
  {S.}~\bibnamefont {Kuhr}},\ }\bibfield  {title} {\enquote {\bibinfo {title}
  {Light-cone-like spreading of correlations in a quantum many-body system},}\
  }\href {\doibase 10.1038/nature10748} {\bibfield  {journal} {\bibinfo
  {journal} {Nature}\ }\textbf {\bibinfo {volume} {481}},\ \bibinfo {pages}
  {484--487} (\bibinfo {year} {2012})}\BibitemShut {NoStop}%
\bibitem [{\citenamefont {Chaudhury}\ \emph {et~al.}(2009)\citenamefont
  {Chaudhury}, \citenamefont {Smith}, \citenamefont {Anderson}, \citenamefont
  {Ghose},\ and\ \citenamefont {Jessen}}]{Nature.461.768}%
  \BibitemOpen
  \bibfield  {author} {\bibinfo {author} {\bibfnamefont {S.}~\bibnamefont
  {Chaudhury}}, \bibinfo {author} {\bibfnamefont {A.}~\bibnamefont {Smith}},
  \bibinfo {author} {\bibfnamefont {B.~E.}\ \bibnamefont {Anderson}}, \bibinfo
  {author} {\bibfnamefont {S.}~\bibnamefont {Ghose}}, \ and\ \bibinfo {author}
  {\bibfnamefont {P.~S.}\ \bibnamefont {Jessen}},\ }\bibfield  {title}
  {\enquote {\bibinfo {title} {Quantum signatures of chaos in a kicked top},}\
  }\href {\doibase 10.1038/nature08396} {\bibfield  {journal} {\bibinfo
  {journal} {Nature}\ }\textbf {\bibinfo {volume} {461}},\ \bibinfo {pages}
  {768--771} (\bibinfo {year} {2009})}\BibitemShut {NoStop}%
\bibitem [{\citenamefont {Parker}\ and\ \citenamefont
  {Chua}(2012)}]{parker2012practical}%
  \BibitemOpen
  \bibfield  {author} {\bibinfo {author} {\bibfnamefont {T.~S.}\ \bibnamefont
  {Parker}}\ and\ \bibinfo {author} {\bibfnamefont {L.}~\bibnamefont {Chua}},\
  }\href@noop {} {\emph {\bibinfo {title} {Practical numerical algorithms for
  chaotic systems}}}\ (\bibinfo  {publisher} {Springer Science \& Business
  Media},\ \bibinfo {year} {2012})\BibitemShut {NoStop}%
\bibitem [{\citenamefont {Ch{\'a}vez-Carlos}\ \emph {et~al.}(2016)\citenamefont
  {Ch{\'a}vez-Carlos}, \citenamefont {Bastarrachea-Magnani}, \citenamefont
  {Lerma-Hern{\'a}ndez},\ and\ \citenamefont {Hirsch}}]{chavez2016classical}%
  \BibitemOpen
  \bibfield  {author} {\bibinfo {author} {\bibfnamefont {J.}~\bibnamefont
  {Ch{\'a}vez-Carlos}}, \bibinfo {author} {\bibfnamefont {M.~A.}\ \bibnamefont
  {Bastarrachea-Magnani}}, \bibinfo {author} {\bibfnamefont {S.}~\bibnamefont
  {Lerma-Hern{\'a}ndez}}, \ and\ \bibinfo {author} {\bibfnamefont {J.~G.}\
  \bibnamefont {Hirsch}},\ }\bibfield  {title} {\enquote {\bibinfo {title}
  {Classical chaos in atom-field systems},}\ }\href {\doibase
  10.1103/PhysRevE.94.022209} {\bibfield  {journal} {\bibinfo  {journal}
  {Physical Review E}\ }\textbf {\bibinfo {volume} {94}},\ \bibinfo {pages}
  {022209} (\bibinfo {year} {2016})}\BibitemShut {NoStop}%
\end{thebibliography}
\end{document}